\documentclass[11pt,letterpaper]{article}

\usepackage[margin=1in]{geometry}

\edef\restoreparindent{\parindent=\the\parindent\relax}
\usepackage{parskip} 
\restoreparindent

\usepackage{amsmath}
\usepackage{amsfonts}
\usepackage{amssymb}
\usepackage[pdftex]{graphicx}
\usepackage{color}
\usepackage{rotating}
\usepackage{bm}
\usepackage{xfrac}

\allowdisplaybreaks

\newtheorem{theo}{Theorem}
\newtheorem{lem}{Lemma}
\newtheorem{ass}{Assumption}
\newtheorem{defin}{Definition}
\newtheorem{prop}{Proposition}
\newtheorem{cor}{Corollary}
\newtheorem{rem}{Remark}
\newtheorem{claim}{Claim}
\newtheorem{obs}{Observation}
\newtheorem{ex}{Example}

\newcommand{\appendixnumbering}{%
\setcounter{theo}{0}\renewcommand{\thetheo}{\Alph{section}\arabic{theo}}%
\setcounter{lem}{0}\renewcommand{\thelem}{\Alph{section}\arabic{lem}}%
\setcounter{prop}{0}\renewcommand{\theprop}{\Alph{section}\arabic{prop}}%
\setcounter{cor}{0}\renewcommand{\thecor}{\Alph{section}\arabic{cor}}%
\setcounter{defin}{0}\renewcommand{\thedefin}{\Alph{section}\arabic{defin}}%
\setcounter{rem}{0}\renewcommand{\therem}{\Alph{section}\arabic{rem}}%
\setcounter{claim}{0}\renewcommand{\theclaim}{\Alph{section}\arabic{claim}}%
\setcounter{obs}{0}\renewcommand{\theobs}{\Alph{section}\arabic{obs}}%
\setcounter{ass}{0}\renewcommand{\theass}{\Alph{section}\arabic{ass}}%
\setcounter{ex}{0}\renewcommand{\theex}{\Alph{section}\arabic{ex}}%
}

\DeclareMathOperator*{\argmax}{arg\,max}

\begin{document}

\title{\textsc{Screening with Discrete Types and Discrete Contracts}\thanks{We thank Pierpaolo Battigalli and anonymous reviewers for detailed feedback on an earlier draft. We also thank seminar participants at UW Bothell and Universidad Torcuato Di Tella. Burkhard gratefully acknowledges financial support through ARO Contract W911NF2210282. This is a companion paper to Francetich and Schipper (2026), where we study rationalizable screening under unawareness.}}

\author{Alejandro Francetich\thanks{School of Business, University of Washington, Bothell. Email: aletich@uw.edu} \and Burkhard C. Schipper\thanks{Department of Economics, University of California, Davis. Email: bcschipper@ucdavis.edu
}}

\date{October 6, 2026}

\maketitle

\begin{abstract}
We study screening when both the agent's types and the contracts available to the principal are finite: Quantities and transfers are drawn from a finite integer grid, while marginal costs lie on a slightly off-set grid, with integer costs as a limit case. We compare two solution concepts, equilibrium and rationalizability. Since rationalizability does not feature an equilibrium tie-breaking convention, we analyze screening under weak and under strict incentives. We develop a discrete first-order approach and characterize the optimal menus for full-support log-concave beliefs. Optimal quantities need not be unique, but at most two adjacent quantities are optimal and multiplicity is a knife-edge case in beliefs. Our rationalizability solution concept, \textit{$\Delta$-O rationalizability}, restricts the principal's beliefs over types and requires her best replies to be robust to belief perturbations. Two rounds of elimination suffice, and the rationalizable and the equilibrium outcomes essentially coincide: The predictions of equilibrium survive without a common prior, while avoiding a tie-breaking convention.
\\


\end{abstract}

\newpage

\section{Introduction} 

In this paper, we consider a principal (she) seeking to procure a good from an agent (he) whose cost structure is his private information. In order to screen the agent, the principal offers a menu of contracts \`a la Mirrlees (1971), Mussa and Rosen (1978), Baron and Myerson (1982), or Maskin and Riley (1984). However, our screening problem is set in an entirely discrete setting: Our agent has finitely many types \emph{and} our principal is restricted to choosing contracts from a \emph{finite grid of quantities and transfers}.

Our interest in discrete contracts is at least two-fold. 
From a practical point of view, they are more realistic: Quantities can only be measured with finite precision, and monetary transfers have a smallest unit---i.e., dollars or cents. While the standard first-order approach to screening is elegant and extremely useful, it is important to scrutinize which issues arise when trying to extend the well-known results of continuous marginal analysis to a discrete setting.

From a theoretical perspective, discreteness facilitates going beyond the standard equilibrium analysis. In settings where the principal and agent have not previously interacted, lack experience in contracting, or face a ``novel'' contracting situation, it cannot be assumed to be common knowledge that the parties abide by equilibrium conventions, or that they even share a prior. The relevant approach here is rationalizability (Bernheim, 1984; Pearce, 1984; Battigalli and Siniscalchi, 2003). Rationalizability notions are also of interest for their interpretability in terms of higher-order beliefs in rationality (Tan and Werlang, 1988; Brandenburger and Dekel, 1987; Battigalli and Siniscalchi, 2002) and finite levels of reasoning (see, for instance, Schipper and Zhou, 2026).

Standard screening analysis rests on one such convention. The principal designs transfers so that the relevant constraints bind, which leaves the agent indifferent; it is then assumed to be common knowledge that he breaks these indifferences in favor of the principal, namely accepting the contract when held to his outside option, and choosing the contract intended for him when (locally) inflating his costs would leave him just as well off. Without this assumption, the principal has to provide strict incentives. Yet in a continuum there is no smallest transfer increase that provides them: Her best replies do not exist, and hence neither do optimal rationalizable menus. This is an artifact of working with a continuum. In reality, a smallest unit does exist, so that strict incentives are feasible.

Working with discrete contracts is the seemingly obvious, multi-purpose fix. However, it introduces its own new subtleties into the analysis. We discuss each in turn. 
The first one has to do with methodology. Using discrete contracts transforms the problem into a discrete optimization problem, so familiar marginal first-order methods do not apply. We develop a discrete first-order approach using discrete strict concavity that closely mimics standard marginal analysis. 

The second one has to do with providing strict incentives. If cost types, quantities, and transfers reside all on the \emph{same} discrete grid, transfers can perfectly match any ``cost + rent'' amount. Incentive and participation constraints hold with equality, resulting once again in (local) indifferences. To provide strict incentives, the principal needs to award each type an ``extra dollar,'' which quickly compounds across types. More importantly, indifferences present an extra challenge under rationalizability; the principal can perfectly rationalize the agent choosing ``wrong'' when indifferent. Our specification of the type grid as non-integer and slightly off-set with respect to the integer contract grid---the grids matching in the limit case---gives us strict incentives for free and reflects the fact that, in reality, quantities and transfers are standardized but costs typically involve a subjective, private component. 

Third, unlike its marginal-analysis counterpart, discrete strict concavity does not imply uniqueness of optimal menus. We show that, for each type, there can be two optimal quantities---but no more than two, and they must be adjacent on the grid. Moreover, nonuniqueness is a knife-edge case that can be resolved by requiring robustness to slight perturbations of the marginal belief. The exception is the case of the most efficient type, whose discrete first-order conditions do not involve the marginal belief at all. We tackle this case through a mild nondegeneracy assumption on the principal's valuation. 

Finally, as in standard marginal analysis, monotone inverse hazard rates imply monotonicity of optimal quantities and allow for constraint reduction. However, even strictly monotone inverse hazard rates can only ensure weak monotonicity; i.e., we can have ``nonbinding'' pooling, and any form of pooling breaks strict incentives. This can happen both when the principal's utility function is ``too concave'' and when the grid of quantities runs out, so we limit its degree of concavity and require the grid to be wide enough that no type is pushed against its upper end. 

We develop optimal discrete strictly concave screening under both weak and strict incentive compatibility and participation constraints and log-concave beliefs over types. We show that the sets of optimal menus under the two formulations coincide whenever cost types are off-set from the integers, and that the optimal menu is unique for almost all such beliefs. With non-integer types, transfers involve rounding up costs. The resulting ``rounding-up rent'' scales with the quantity, yet it stays below a full unit. Thus, no type is ever indifferent between two distinct contracts. Along the way, we uncover a feature of the discrete problem that has no counterpart in the continuum: When the principal is indifferent between two adjacent optimal quantities for a given type, the agent is not. He strictly prefers the contract with the larger quantity, since it leaves him the larger rent.

We then apply our analysis to equilibrium screening in a purely discrete setting. Note that with the equilibrium screening approach we do not need to require strict incentive compatibility or strict participation constraints, types to be off-set from integers, a limit on discrete strict concavity, or that the least efficient type be served. Yet, what equilibrium does require is a common prior over types and the convention that the agent breaks indifferences in favor of the principal.

More importantly, we develop a rationalizability approach to screening. Our solution concept, \textit{$\Delta$-O rationalizability}, restricts the principal's marginal beliefs over the agent's types to be full-support, strictly log-concave (which implies single-peakedness), and such that even the least efficient type is worth serving. Within these restrictions, we require robustness: A menu must be optimal not just w.r.t. one such marginal belief, but w.r.t. an open set of such marginal beliefs. We show that with types off-set from the integers, we can solve the discrete screening game in two rounds of elimination. Moreover, every rationalizable menu corresponds to an equilibrium menu for an open set of common priors, and vice versa. Essentially, the set of equilibrium menus and the set of rationalizable menus coincide. In other words, the predictions of equilibrium screening survive without a common prior, while avoiding a tie-breaking convention.

Our analysis serves as the full-awareness benchmark for a companion paper on screening under unawareness, Francetich and Schipper (2026), where the principal is initially \textit{unaware} of some of the agent's types. The agent has the option to raise the principal's awareness, totally or partially, before contracting. Further restricting beliefs to satisfy reverse Bayesianism and a ``wariness'' condition, we show that if the principal is ex ante only unaware of high-cost types, all of these types have an incentive to raise her awareness---otherwise, they would not be served. With three types, the two lower types also want to alert the principal: The presence of an even higher type makes the original two look better. But with more than three types, this intuition may break down if the principal focuses her ``revised'' beliefs too much on the higher-cost types. When the principal is ex ante only unaware of more efficient (i.e., lower-cost) types, then \textit{no type} raises her awareness, leaving her none the wiser. Intuitively, raising awareness of lower-cost types makes all of them look worse in the eyes of the principal, while said types can enjoy a higher surplus under the ``default'' offer. This analysis is done with an extension of our present rationalizability concept to unawareness. Settings under unawareness often model novel situations in which equilibrium conventions could not have been learned. 

Our paper contributes to the classic literature on screening. Both the standard case with a continuum of types and the case of finite types and a continuum of quantities and transfers have been extensively studied and applied throughout economics (see, for instance, Laffont and Martimort, 2001). Yet, we are not aware of any study using discrete contracts, even though they are very realistic. 

We also contribute to the literature on rationalizability in games with incomplete information (Battigalli, 2003, 2006; Battigalli and Siniscalchi, 2003; Heifetz, Meier, and Schipper, 2021; Li and Schipper, 2020). Rationalizability with belief restrictions has been studied at least since Battigalli and Siniscalchi (2003) and Battigalli (2003) under the name of $\Delta$-Rationalizability (see also Battigalli and Friedenberg, 2012; Battigalli and Prestipino, 2013; Brandenburger, Friedenberg, and Keisler, 2008; Brandenburger and Friedenberg, 2010). However, we are not aware of any rationalizability concept with the same restrictions as ours. The closest approach---in static games with complete information---is by Ziegler and Zuazo-Garin (2020), who require a rationalizable strategy to be a best response to all beliefs in a subset of beliefs over remaining strategies. Their aim is to model cautiousness and ambiguity, while we use the idea in the context of incomplete information to model robustness to small perturbations of beliefs. 
 
The rest of the paper is organized as follows. Section~\ref{sec: discrete_concave_screening} presents the model. Section~\ref{sec: screening} analyzes the principal's screening problem, under both the weak and the strict formulations of the incentive-compatibility and participation constraints. Section~\ref{sec: equilibrium} analyzes equilibrium screening, where there is a common prior on types and where the incentives provided are weak, assuming that the agent resolves his indifference in favor of the principal. Section~\ref{sec: rationalizability} develops our robust rationalizability analysis, focusing on strict incentives. Section~\ref{sec: discussion} relates the rationalizable outcomes to the equilibrium ones, summarizes in Table~\ref{assumptions_table} which assumptions and conditions each of the two analyses requires, discusses extensions to nonlinear cost structures, and asks what remains of the rationalizability analysis when only weak incentives are imposed, in which case the agent can be rationalized as (locally) over-reporting his costs. Appendix~\ref{discrete_c_o} presents tools and results of discrete concave optimization. Appendix~\ref{const_simp} contains the constraint-simplification results, while Appendix~\ref{others} collects implications of log-concavity of beliefs over types. Finally, Appendix~\ref{non_linear_app} elaborates on the nonlinear cost extension, where monotonicity of the quantities fails even under strictly increasing virtual costs, and where we have an \textit{upward} distortion at the top, which cannot happen in the linear-cost benchmark.


\section{Model}\label{sec: discrete_concave_screening}

A principal ($P$, ``she'') wants to procure a quantity $q$ of an object from an agent  ($A$, ``he'') in exchange for a transfer $t$. We assume that $q, t \in D := \{0, 1, \ldots, b\}$ for some $b \in \mathbb{N}$. The agent's marginal cost is his private information; we denote it by $\theta \in \Theta$, where:
$$\Theta = \left\{i - \frac{1}{\gamma}: i = 1, \ldots, m\right\}$$
for $\gamma \in \mathbb{N} \cup \{\infty\}$, $m \in \mathbb{N}$, and $\gamma > b > m$. We require that both $\gamma$ and $b$ are large compared to $m$, so that the principal is able to design contract menus with as many as $m$ contracts if she desires to do so.\footnote{Although our representation of cost types is unidimensional, we can think of $\theta \in \Theta$ as a score that aggregates the impact of various factors affecting marginal costs, reminiscent of ``pseudotypes'' in scoring auctions; see, for instance, Asker and Cantillon (2008) or Bajari, Houghton, and Tadelis (2014).} 

The case $\gamma < \infty$ represents the case of non-integer types, reflecting the fact that costs can be subjective and more idiosyncratic while contract terms are more objective or standardized; the limit case $\gamma=\infty$ is simply the case of integer types. Our specification further gives us the flexibility to easily accommodate both weak and strict incentives, and allows us to replicate the usual structure of optimal transfers (Section~\ref{sec: screening}).

The principal screens the agent by offering him the choice of a contract $\bm{c} := (q, t)$ from a menu of contracts $M$. The set of contract menus is $\mathcal{M} := 2^{D^2} \setminus \{\emptyset\}$. Presented with a menu $M \in \mathcal{M}$, the agent chooses either a contract $\bm{c}\in M$ or takes his outside option, $\bm{o}$, identified by the null-contract $(0, 0)$, and the game ends. The payoff for the agent with type $\theta$ from signing contract $\bm{c}$ is given by $u_A(\bm{c}, \theta) := t - \theta q$, while he earns $u_A(\bm{o}, \theta)=0$ from the outside option. The principal's utility of output is given by a function $v: D \longrightarrow \mathbb{R}$, and her net utility or payoff from a contract $\bm{c} = (q, t)$ is $u_P(\bm{c}) = v(q) - t$. We assume the following about $v(q)$.

\begin{ass}\label{properties_v_basic} The function $v$ satisfies the following properties:
\begin{enumerate}
\item\label{normalization}
$v(0) = 0$;
\item\label{sincreasing}
$v$ is strictly increasing in $q$;
\item\label{sdc}
$v$ is strictly discrete concave in $q$: For every $q \in \{1, \ldots, b - 1\}$, $v(q + 1) + v(q - 1) < 2 v(q)$.
\end{enumerate}
\end{ass} 

Assumptions~\ref{properties_v_basic}.\ref{normalization} and~\ref{properties_v_basic}.\ref{sincreasing} are standard in the screening literature, while Assumption~\ref{properties_v_basic}.\ref{sdc} is the discrete counterpart of the typical assumption of strict concavity. We can characterize the second and third conditions in terms of \textit{discrete derivatives}. Define $\Delta^- v(q) := v(q) - v(q - 1)$ for $q \in \{1, \ldots, b\}$ and $\Delta^+ v(q) := v(q + 1) - v(q)$ for $q \in \{0, \ldots, b - 1\}$; these are the discrete backward and forward derivatives of $v(q)$, respectively.\footnote{At the two ends of the grid, we adopt the conventions $\Delta^{-} v(0) := + \infty$ and $\Delta^{+} v(b) := - \infty$, so that first-order conditions remain meaningful at $q = 0$ and at $q = b$.} Assumption~\ref{properties_v_basic}.\ref{sincreasing} holds if and only if $\Delta^{-}v(q)>0$ for all $q$, or equivalently $\Delta^{+}v(q)>0$ for all $q$ (see Observation~\ref{d_monot} in the appendix). Defining the second discrete derivative $\Delta^+ \Delta^- v(q) := \Delta^+(\Delta^- v(q))$, we can also restate Assumption~\ref{properties_v_basic}.\ref{sdc} as $\Delta^+ \Delta^- v(q) < 0$. 

The screening game proceeds in three stages: At the first stage, nature selects the payoff type of the agent, $\theta \in \Theta$. The payoff type is private information of the agent and not known to the principal. That is why at a second stage the principal offers a menu of contracts $M \in \mathcal{M}$ to the agent, who in the third and final stage selects a contract $\bm{c} = (q, t) \in M$ from the menu or his outside option $\bm{o}$. We will consider both equilibrium and rationalizable outcomes. To this end, we first study the principal's constrained optimization problem, both under weak and strict inequality constraints. For these optimization problems, the principal forms a full-support belief about types, $p \in \text{int}(\Delta(\Theta))$, which represents the common prior in equilibrium screening (Section~\ref{sec: equilibrium}) and the marginal over types of the principal's joint beliefs over the agent's behavior and types under rationalizability (Section~\ref{sec: rationalizability}).
 
To facilitate the analysis, define $\theta_{i} := i - \sfrac{1}{\gamma}$ for $i = 1, \ldots, m$. Moreover, denote by $(q_i,t_{i})$ the contract designed for type $\theta_{i}$ and by $p_i$ the probability of $\theta_{i}$ under $p$.


\section{Weak Versus Strict Constraints\label{sec: screening}}

In this section, we consider two formulations of the constrained optimization problem at the heart of solving the screening game: weak vs. strict incentive-compatibility and participation constraints. As discussed in the introduction, the latter is important for deriving rationalizable outcomes of the screening game while the former is relevant for equilibrium. For both formulations, we demonstrate how constraint simplification works in our entirely discrete setting and under our type specification. 

\subsection{Constraint Simplification and Discrete F.O.C.s}

The \emph{discrete screening problems} are given as follows: For both formulations, the principal maximizes the expected profit function:
$$\max_{(q_i, t_{i})_{i = 1,\ldots, m} \in (D^2)^m} \sum_{i = 1}^{m} p_i \left(v(q_i) - t_{i}\right).$$ 
The two screening versions are distinguished by weak versus strict incentive compatibility and participation constraints: For all $i, j=1,\ldots,m$,

$$\begin{array}{l|l} \mbox{Weak inequality constraints} & \mbox{Strict inequality constraints} \\ \hline 
\mbox{IC}_{i, j}: \quad t_{i} - \theta_{i}q_i \geq t_{j} - \theta_{i}q_{j} \quad & \mbox{SIC}_{i, j}: \quad t_{i} - \theta_{i}q_i > t_{j} - \theta_{i}q_{j} \\
\mbox{PC}_i: \quad t_{i} - \theta_{i} q_i \geq 0 \quad & \mbox{SPC}_i: \quad t_{i} - \theta_{i} q_i > 0
\end{array}$$ 
Notice that a type awarded no output, $q_i = 0$, satisfies SPC$_i$ only if he is paid a strictly positive transfer for nothing. We will rule out shutdown below.

While many standard constraint-simplification results extend immediately to a discrete setting, there is one potential problem in reducing the number of (local) incentive-compatibility constraints. The usual argument for ruling out downward constraints (i.e., not understating his cost type) relies on the upward ones (i.e., not inflating his cost type) binding. But discreteness changes the sense in which constraints ``bind.'' Under our type-space specification, we verify in Appendix~\ref{const_simp} that we can reduce the constraints as ``usual'': Optimal quantities must satisfy monotonicity (i.e., higher-cost types cannot be awarded a higher quantity than lower-cost types), strictly so under strict inequality constraints; under monotonicity, local incentive-compatibility constraints imply global ones; the only relevant incentive-compatibility constraints are the (local) upward ones; and the only relevant participation constraint is that of $\theta_{m}$, the least-efficient type. The (local) downward constraints are not dropped but shown to be slack at the optimal transfers when $\gamma \neq \infty$; for $\gamma = \infty$ they hold only when consecutive quantities differ by at least two units (Proposition~\ref{local_SIC}). 

The \emph{reduced discrete screening problems} are given as follows: For both problems, the principal maximizes the expected profit function:
$$\max_{(q_i, t_{i})_{i = 1,\ldots, m} \in (D^2)^m} \sum_{i = 1}^{m} p_i \left(v(q_i) - t_{i}\right).$$ subject to: For all $i = 1, \ldots, m - 1$,

$$\begin{array}{l|l} \mbox{Reduced weak inequality constraints} & \mbox{Reduced strict inequality constraints} \\ \hline 
\mbox{IC}_{i, i + 1}: \quad t_{i} - \theta_{i}q_i \geq t_{i + 1} - \theta_{i}q_{i + 1} \quad & \mbox{SIC}_{i, i + 1}: \quad t_{i} - \theta_{i}q_i > t_{i + 1} - \theta_{i}q_{i + 1} \\
\mbox{PC}_m: \quad t_{m} - \theta_{m} q_{m} \geq 0 \quad & \mbox{SPC}_m: \quad t_{m} - \theta_{m} q_{m} > 0 \\
\mbox{M}_{i, i + 1}: \quad q_i \geq q_{i + 1} \quad & \mbox{SM}_{i, i + 1}: \quad q_i > q_{i + 1} 
\end{array}$$

In the optimal contract, transfers are set so that the relevant constraints ``bind.'' However, binding looks a bit different here, depending on how the cost grid and the transfer grid compare. With non-integer costs and integer transfers, the optimal feasible transfers involve rounding up and thus ensure strict incentives for free. When $\gamma = \infty$, namely with integer costs, providing strict incentives requires adding an ``extra dollar'' to the usual ``cost $+$ information rent'' terms. A convenient way to accommodate both cases is to add the term $1 - \left\lceil \sfrac{1}{\gamma}\right\rceil$ to the transfers, a term that is zero for every finite $\gamma$ and equal to 1 when $\gamma = \infty$. 

Denoting by $\mathbb{I}_s$ the indicator function for strict incentives ($\mathbb{I}_s=1$ when we deal with the strict problem and $\mathbb{I}_s=0$ otherwise), optimal transfers become:
\begin{align*} 
t_{m} & = \left\lceil \theta_{m} \cdot q_{m} \right\rceil + \mathbb{I}_s\left(1 - \left\lceil \frac{1}{\gamma}\right\rceil\right),
\end{align*} 
and for $i = 1, \ldots, m - 1$,
\begin{align*}
t_{i} & =\left\lceil \theta_{i} q_i +t_{i + 1} - \theta_{i} q_{i + 1} \right\rceil + \mathbb{I}_s\left(1 - \left\lceil \frac{1}{\gamma}\right\rceil\right) = t_{i + 1}+\left\lceil\theta_{i}\left(q_i - q_{i + 1}\right) \right\rceil + \mathbb{I}_s\left(1 - \left\lceil \frac{1}{\gamma}\right\rceil\right).
\end{align*} 

Now, it is not true in general that $\lceil \alpha n\rceil = \lceil \alpha\rceil n$ for any $\alpha>0$ and integer $n$; for instance,  $\lceil 2.5 \times 2 \rceil = 5 < 6 = \lceil 2.5\rceil \times 2$. But it is true under our type specification (see Claim~\ref{product} in Appendix~\ref{const_simp}). So, noting that $\lceil\theta_{j}\rceil = j$ for $j=1,\ldots,m$, we show in Proposition~\ref{binding} in Appendix~\ref{const_simp} that optimal transfers can be rewritten as:
\begin{align}
t_{m} & = m q_{m} + \mathbb{I}_s\left(1 - \left\lceil \frac{1}{\gamma}\right\rceil\right),\label{optimal_ts_m} \\
t_{i} & = i q_i + \sum_{j = i + 1}^{m} q_{j} + (m - i + 1) \mathbb{I}_s \left(1 - \left\lceil \frac{1}{\gamma}\right\rceil\right).\label{optimal_ts_i}
\end{align} 
This formulation allows us to replicate the usual structure of transfers as ``cost $+$ information rent,'' with the cost rounded up plus any ``extra dollar'' transfers in the case of integer types and strict constraints. For every $i=1,\ldots,m$, we can write the cost term as: $$i q_i=\theta_{i}q_{i}+\frac{1}{\gamma}q_{i};$$ we call the second term on the right-hand side the agent's ``rounding-up'' rent. Notice that this rent scales with the quantity.

When temporarily omitting the (strict) monotonicity constraints (S)M$_{i, i+1}$, the formulation of transfers allows us to transform the constrained discrete optimization problems into the following unconstrained discrete optimization problem:  
\begin{align*} \max_{(q_i)_{i = 1,\ldots, m} \in D^m} \sum_{i = 1}^m p_i \left(v(q_i) - i q_i - \sum_{j = i + 1}^{m} q_{j}\right) - \left(\sum_{i = 1}^m p_i(m - i + 1)\right) \mathbb{I}_s\left(1 - \left\lceil \frac{1}{\gamma}\right\rceil\right).
\end{align*} 
By the ``reshuffling'' of sums, $$\sum_{i = 1}^{m} p_i \sum_{j =i+1}^{m} q_{j} = \sum_{i = 1}^m \left(\sum_{j< i} p_j q_i\right) = \sum_{i = 1}^m p_i \left(\frac{\sum_{j<i}p_j}{p_i} q_i\right);$$ with the normalization $\sum_{j < 1} p_j \equiv 0$, the problem can be rewritten as: 
\begin{align} 
\max_{(q_i)_{i = 1,\ldots, m} \in D^m} \sum_{i = 1}^m p_i \left(v(q_i) - \left(i+\frac{\sum_{j<i} p_j}{p_i}\right) q_i\right) - \left(\sum_{i = 1}^m p_i(m - i + 1)\right) \mathbb{I}_s\left(1 - \left\lceil \frac{1}{\gamma}\right\rceil\right).\label{objective}
\end{align}

Maximization of~\eqref{objective} is done term by term, and the last term does not depend on quantities. Thus, for both weak and strict formulations of the discrete screening problem, the principal wants to maximize $v(q_i) - \varphi_{i}(p) q_i$ for each $i = 1,\ldots , m$, where: 
$$\varphi_i(p) := i + \frac{\sum_{j < i} p_j}{p_i}$$ 
is the discrete version of the virtual cost of the rounded-up type $\theta_{i}$ given $p$.

\begin{figure}[h!]
\medskip
\centering
\includegraphics[width=.49\linewidth]{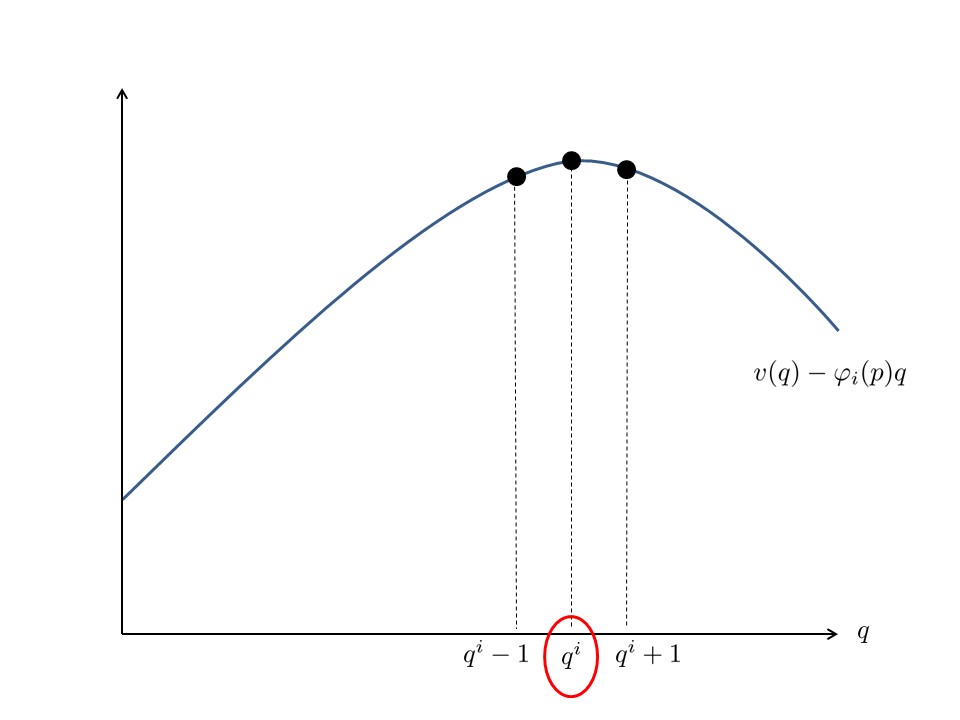}\hfill
\includegraphics[width=.49\linewidth]{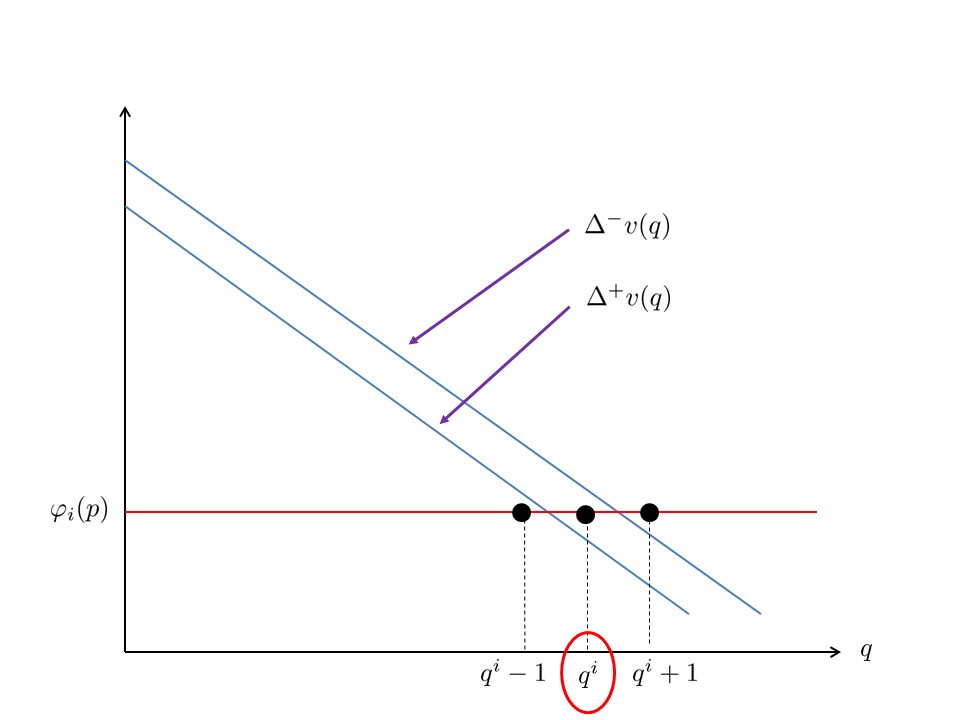}
\caption{F.O.C.s for the reduced principal's problem for type $\theta_{i}$.\label{maximizing}}
\medskip
\end{figure}

Employing the tools of discrete concave optimization presented in Appendix~\ref{discrete_c_o}, we derive the discrete analogues of the calculus-based first-order conditions for $i = 1, \ldots, m$:
\begin{eqnarray}\label{combined_FOCs}  
\Delta^+ v(q_i) \leq \varphi_{i}(p) \leq \Delta^- v(q_i).
\end{eqnarray} 
See Figure~\ref{maximizing}. The left panel of Figure~\ref{maximizing} depicts schematically the term corresponding to type $\theta_{i}$ in the principal's payoff function~\eqref{objective} as well as the corresponding optimal quantity; the right panel shows how the discrete derivatives of the principal's utility function ``sandwich'' the optimal quantity at the corresponding virtual cost. 

As usual, each type's quantity equates the principal's marginal value to that type's \textit{virtual} cost rather than to his true cost. The wedge $\sum_{j < i} p_j/p_i$ vanishes for the most efficient type and is positive for all others, whose quantities are therefore distorted downward; this is how the principal saves on information rents. However, for $\gamma < \infty$, virtual costs entail rounding up. So, we have potentially an additional source of downward distortion: $$\varphi_{i}(p)>\theta_{i}+\frac{\sum_{j < i} p_j}{p_i}.$$ In particular, the lowest type may be commissioned an amount of output that falls short of the efficient one. Nonetheless, this possible extra distortion vanishes when $\gamma = \infty$---where we recover the usual ``no distortion at the top'' result. 

Of course, for the solutions to~\eqref{combined_FOCs} to constitute a valid contract menu, they must satisfy the corresponding monotonicity constraints. Before addressing this point, we illustrate the principal's reduced problem and the conditions in~\eqref{combined_FOCs} in an example. 

\begin{ex}\label{example_1}
Let $D = \{0, 1, \ldots, 50\}$, $\Theta = \{0.99, 1.99, 2.99\}$ (here, $\gamma = 100$), and $v(q) = 50 q - 0.5 q^{2}$; we have $\Delta^{+} v(q) = 49.5 - q$ and $\Delta^{-} v(q) = 50.5 - q$. Consider the marginal belief given by $p = (p_1, p_2, p_3) = (\frac{1}{3}, \frac{1}{3}, \frac{1}{3})$, which yields $\varphi_{1}(p) = 1$, $\varphi_{2}(p) = 3$, and $\varphi_{3}(p) = 5$. The F.O.C. for $q_{1}$ is:
\[
49.5-q_{1} \leq 1 \leq 50.5 - q_{1};
\] 
rearranging terms leads to $48.5 \leq q_{1} \leq 49.5$, which yields $q_{1} = 49$. For $q_ {2}$,
\begin{align*}
&49.5-q_{2} \leq 3 \leq 50.5 - q_{2}.
\end{align*}
Now, we get $46.5 \leq q_{2}\leq 47.5$, so $q_{2} = 47$. Finally, for $q_{3}$,
\begin{align*}
&49.5 - q_{3} \leq 5 \leq 50.5 - q_{3};
\end{align*} 
we can rewrite the inequalities as $44.5 \leq q_{3} \leq 45.5$, so we get $q_{3} = 45$. \hfill $\Box$
\end{ex}

\subsection{Implementability: Monotonicity and (Non-)Uniqueness}

Before addressing the omitted monotonicity constraints, a note on multiplicity. One notable difference with the case of continuum contracts is that, even under strict concavity of $v$, the F.O.C.s need not have a unique solution. We illustrate nonuniqueness in the following example. 

\begin{ex}\label{example_nonuniqueness}
Consider the same setting as in Example~\ref{example_1}, but now let the marginal belief be given by $p_1 = p_3 = 0.25$, and $p_2 = 0.5$. Now, $\varphi_{1}(p) = 1$, $\varphi_{2}(p) = 2.5$, and $\varphi_{3}(p) = 6$. The F.O.C. for $q_{1}$ is independent of the beliefs, so we once again get $q_{1} = 49$. For $q_{3}$, we now have:
\begin{align*}
&49.5 - q_{3} \leq 6 \leq 50.5 - q_{3};
\end{align*} 
we can rewrite the inequalities as $43.5 \leq q_{3} \leq 44.5$, so we get $q_{3} = 44$. Finally, for $q_{2}$,
\begin{align*}
&49.5 - q_{2} \leq 2.5 \leq 50.5 - q_{2}.
\end{align*}
Now, we get $47 \leq q_{2} \leq 48$, so we have two possible solutions: $q_{2} = 47$ and $q_{2} = 48$.  \hfill $\Box$
\end{ex}

In fact, for any type, there can be two solutions; but no more than two, and they must be adjacent. This holds both for the formulation with weak and strict inequality constraints, as it is based on the F.O.C.s for quantities. Moreover, $q_i$ and $q_i + 1$ are the two solutions for type $\theta_{i}$ if and only if $\Delta^- v(q_i + 1) = \Delta^{+} v(q_i) = \varphi_{i}(p)$. Proposition~\ref{nonuniqueness_gen} in the appendix provides a general proof, but we present these statements in the context of our principal's problem below.

\begin{prop}[Multiplicity]\label{nonuniqueness} For each $i = 1, \ldots, m$, there are at most two solutions to~\eqref{combined_FOCs}. If $q_i$ and $q'_i$ are solutions, then either $q_i = q'_i$ or $\max\{q_i, q'_i\} = \min\{q_i, q'_i\} + 1$. The latter case happens if and only if $\Delta^- v(\max\{q_i,q'_{i}\}) = \varphi_{i}(p) = \Delta^+ v(\min\{q_i,q'_{i}\})$.  
\end{prop}

\begin{figure}[h!]
\medskip
\centering
\includegraphics[width=.49\linewidth]{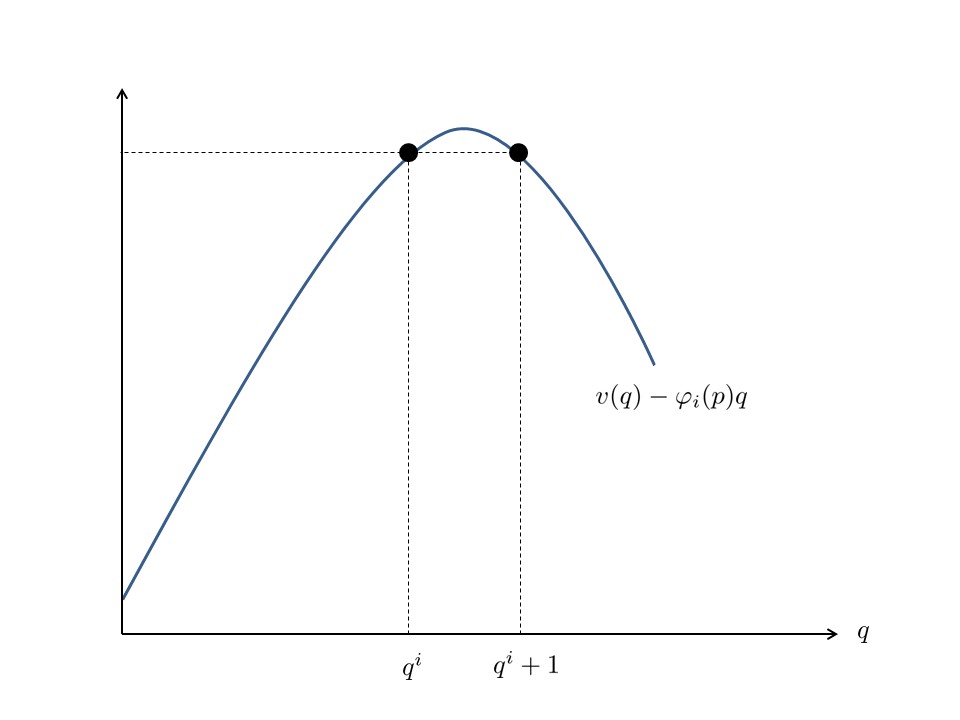}\hfill
\includegraphics[width=.49\linewidth]{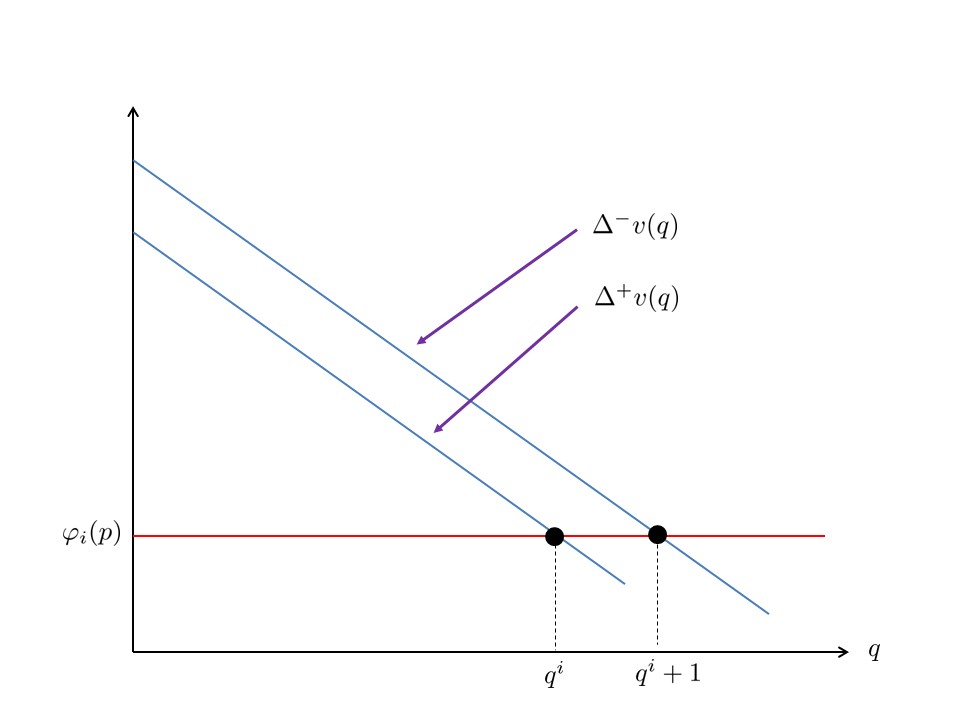}
\caption{There are two different optimal quantities for type $\theta_i$.\label{non-unique}}
\medskip
\end{figure}

Figure~\ref{non-unique} illustrates the condition identified in Proposition~\ref{nonuniqueness}. The two optimal quantities attain the exact same value of $v(q) - \varphi_{i}(p) q$. If quantities could be chosen from a continuum and $v$ were extended accordingly, all intermediate values for $q$ would be even better; however, such intermediate values are infeasible under discrete contracts. 

We now put multiplicity aside until the end of the section, and take up the ignored monotonicity constraints. With finite types, a commonly assumed sufficient condition for the monotonicity constraints not to bind is log-concavity of $p$ (Bagnoli and Bergstrom, 2005).

\begin{defin}[Log-concavity]\label{log_conc} Probability distribution $p \in \Delta(\Theta)$ is \textit{log-concave} if for all $i = 2, \ldots, m - 1$, $p_i p_i \geq p_{i + 1} p_{i - 1}$, and \textit{strictly log-concave} if all these inequalities are strict.
\end{defin}

Equivalently, $\Delta^{+} \Delta^{-} \log p_{i} = \log p_{i + 1} - 2 \log p_{i} + \log p_{i - 1} = \log \left(\frac{p_{i + 1} p_{i - 1}}{p_{i}^{2}}\right)$, so $p$ is log-concave if and only if $\Delta^{+} \Delta^{-} \log p_{i} \leq 0$ for $i = 2, \ldots, m - 1$, and strictly log-concave if and only if these inequalities are strict. Log-concavity of $p$ is thus discrete concavity of $\log p$ in the sense of Appendix~\ref{discrete_c_o}.

We denote the set of full-support, log-concave beliefs as:
$$\Delta^{FL}(\Theta) := \left\{p \in \text{int}(\Delta(\Theta)) : p \mbox{ is log-concave}\right\}.$$
Log-concavity leads to a nonincreasing likelihood ratio (see Proposition~\ref{hazard_rates}): $$\frac{p_{i+k}}{p_i} \geq \frac{p_{j+k}}{p_j}$$ for any $i, j = 1, \ldots, m$ and $k$ such that $j > i$ and $j + k \leq m$; and to a strictly increasing discrete inverse hazard rate or \textit{Mills' ratio} (Mills, 1926): For any $i, j = 1, \ldots, m$ with $j > i$, 
$$\frac{\sum_{k < j} p_k}{p_j} > \frac{\sum_{k < i} p_k}{p_i}.$$
It can also be shown that log-concavity yields virtual costs that increase by more than one unit from one type to the next: $\varphi_{i+1}(p) - \varphi_i(p) > 1$ for all $i = 1,\ldots, m-1$ (see Corollary~\ref{monotonicity_virtual_valuations}), so in particular the distribution is regular in the Myerson sense: the ranking of types by virtual costs matches the ranking of ``raw'' types.

\begin{prop}[Weak monotonicity]\label{monotonicity_weak_suf} Under Assumption~\ref{properties_v_basic} and for any $p \in \Delta^{FL}(\Theta)$, the solutions to~\eqref{combined_FOCs} satisfy $\mbox{M}_{i, i + 1}$ for all $i = 1, \ldots, m - 1$. If for some types $i, i + 1$ we have two solutions $q_i, \hat{q}_{i}$ and $q_{i + 1}, \hat{q}_{i+1}$, then $\min\{q_i, \hat{q}_{i} \} \geq \max\{q_{i + 1}, \hat{q}_{i+1}\}$. 
\end{prop}

\noindent \textsc{Proof. } We start with the case where all optimal quantities are unique. For $i = 1, \ldots, m - 1$, let $q_i$ and $q_{i + 1}$ be the solutions to the corresponding F.O.C.s, $\Delta^{+} v(q_i) \leq \varphi_{i}(p) \leq \Delta^{-} v(q_i)$ and $\Delta^{+} v(q_{i + 1}) \leq \varphi_{i+1}(p) \leq \Delta^{-} v(q_{i + 1})$. Assume that we have some $i = 1, \ldots, m - 1$ such that $q_i < q_{i + 1}$. Then, by Lemma~\ref{monotoncity_derivative}(iv) in the appendix, $\Delta^{-} v(q_{i + 1}) \leq \Delta^{+} v(q_i)$. Combining the conditions, we get:
\begin{align*}
\varphi_{i+1}(p) \leq \Delta^{-} v(q_{i + 1}) \leq \Delta^{+} v(q_i) \leq \varphi_{i}(p),
\end{align*}
which contradicts Corollary~\ref{monotonicity_virtual_valuations}, an implication of log-concavity. 

If there are two optimal quantities for either type $i$ and/or $i+1$, $q_i, \hat{q}_{i}$ and $q_{i + 1}, \hat{q}_{i+1}$, repeat the argument for $\tilde{q}_{i} = \min\{q_i,\hat{q}_{i}\}$ and $\tilde{q}_{i+1}=\max\{q_{i + 1}, \hat{q}_{i+1}\}$. \hfill $\Box$\\

Log-concavity is sufficient for weak monotonicity of the quantities $\mbox{M}_{i, i + 1}$, which is all we ask for when we impose weak incentives. However, it does not guarantee strict monotonicity $\mbox{SM}_{i, i + 1}$ even if the discrete Mills' ratio is strictly monotone---rendering it insufficient for the reduction of strict inequality constraints. The next example features a strictly log-concave type distribution that yields a fully pooled menu.

\begin{ex}\label{weak_mono} Consider $D = \{0, 1, \ldots, 51\}$, the utility function $v(q) = 253q - 2.5q^{2}$, types $\Theta=\{0.99, 1.99, 2.99\}$ (again $\gamma = 100$). Here, $\Delta^{+} v(q) = 250.5 - 5q$, $\Delta^{-} v(q) = 255.5 - 5q$. Consider the marginal belief $p_{1}=\frac{2}{9}$, $p_{2}=\frac{4}{9}$, and $p_{3} = \frac{1}{3}$. This belief is strictly log-concave: $p_{2}^2=\frac{16}{81}>\frac{6}{81}=\frac{2}{27}=p_{1}p_{3}$. We have $\varphi_{1}(p) = 1$, $\varphi_{2}(p) = 2.5$, and $\varphi_{3}(p) = 5$. Optimal quantities are:
\begin{align*}
& 250.5 - 5 q_{1} \leq 1 \leq 255.5 - 5 q_{1} \ \Rightarrow \ q_{1} = 50;
\\
& 250.5 - 5 q_{2} \leq 2.5 \leq 255.5 - 5 q_{2} \ \Rightarrow \ q_{2} = 50;
\\
&250.5 - 5 q_{3} \leq 5 \leq 255.5 - 5 q_{3} \ \Rightarrow \ q_{3} = 50.
\end{align*}
All types are ``naturally'' bunched: The monotonicity constraints hold with equality without binding (i.e., without ironing). \hfill $\Box$
\end{ex}

\begin{figure}[h!]
\medskip
\centering
\includegraphics[width=.49\linewidth]{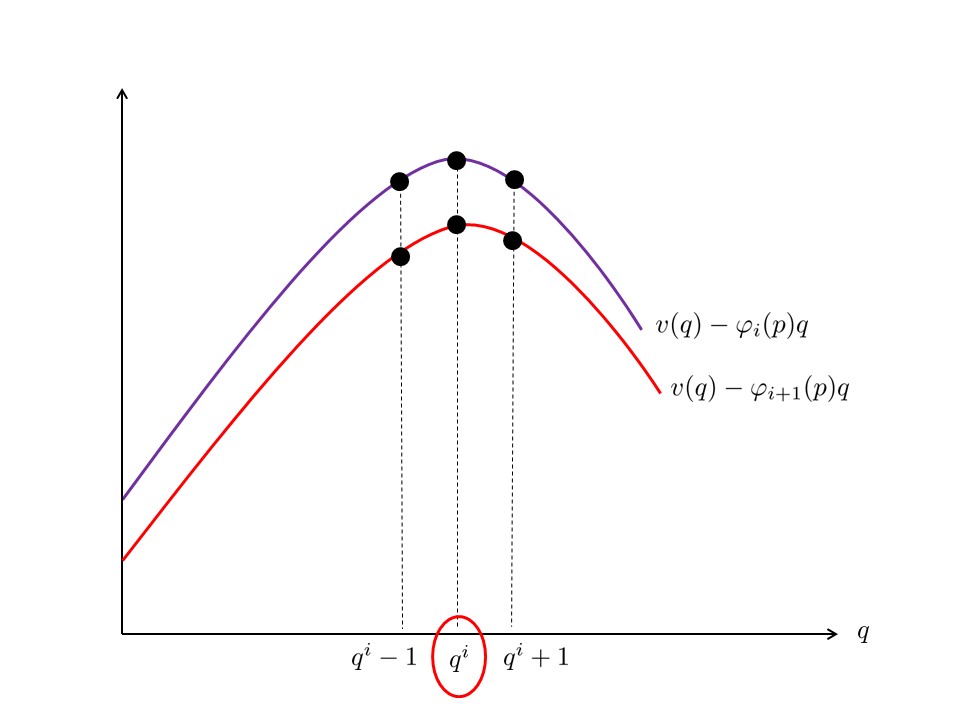}\hfill
\includegraphics[width=.49\linewidth]{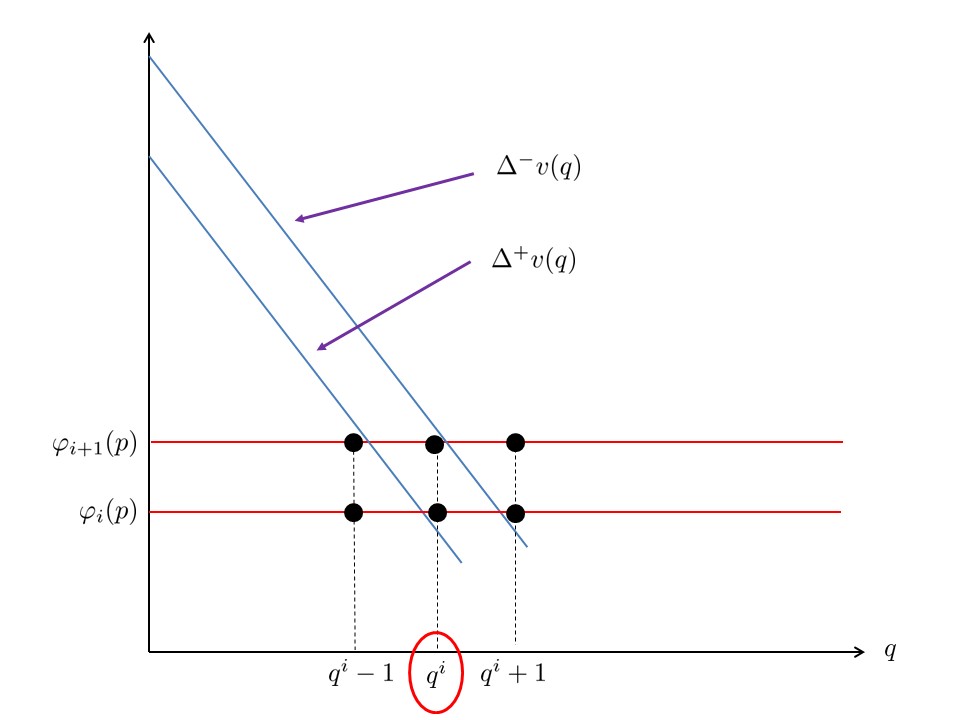}
\caption{Types $\theta_i$ and $\theta_{i + 1}$ are awarded the same quantity.\label{weak_mono_g}}
\medskip
\end{figure}

No amount of log-concavity of the type distribution rules this out; what makes Example~\ref{weak_mono} possible is that $v$ is ``too concave.'' Writing $\Delta^{+} v(q) = A - c q$ and $\Delta^{-} v(q) = A + c - c q$ with $c = \left|\Delta^{+} \Delta^{-} v(q)\right| = 5$, the F.O.C.s place $q_i$ in an interval of length $1$ whose endpoints are $(A - \varphi_{i}(p))/c$ and $(A - \varphi_{i}(p))/c + 1$. Consecutive types' intervals are thus shifted by only $(\varphi_{i + 1}(p) - \varphi_{i}(p))/c$---equal to $\sfrac{3}{10}$ for $i = 1$ and to $\sfrac{1}{2}$ for $i = 2$, both less than $1$---so that they can ``trap'' the same integer, as they do here; see Figure~\ref{weak_mono_g}. What matters here is the size of $c$ relative to the \emph{differences} between consecutive virtual costs, not relative to their levels. 

Pooling is incompatible with the strict incentive-compatibility constraints: When $\gamma \neq \infty$, the term $1 - \left\lceil \sfrac{1}{\gamma}\right\rceil$ vanishes, so bunched types receive the same contract and SIC$_{i, i+1}$ compares a contract with itself; when $\gamma = \infty$, the extra-dollar terms $(m - i + 1)$ in equation~\eqref{optimal_ts_i} scale up with reported efficiency, so with a common quantity every type would rather claim to be $\theta_{1}$.

To ensure strict monotonicity, we restrict the ``degree of concavity'' of $v$ by setting a lower bound on $\Delta^+ \Delta^- v(q)$.\footnote{Recall that Assumption~\ref{properties_v_basic}.\ref{sdc} can be written as $0 > \Delta^+ \Delta^- v(q)$.} Assumption~\ref{not_too_sdc} below rules out such pooling: It ensures that $\Delta^{-} v(q) - \Delta^{+} v(q) \leq 1$ at every $q$, while Corollary~\ref{monotonicity_virtual_valuations} gives $\varphi_{i + 1}(p) - \varphi_{i}(p) > 1$, so consecutive types cannot be awarded the same quantity. In Example~\ref{weak_mono} this difference equals $c = 5$, which is why the intervals in Figure~\ref{weak_mono_g} trap the same integer.

\begin{ass}\label{not_too_sdc} For every $q \in \{1, \ldots, b - 1\}$, $\Delta^+ \Delta^- v(q) \geq -1$.
\end{ass} 

Strict monotonicity also requires the grid of quantities to be wide enough that no type is pushed against its upper end: if two consecutive types were both awarded the maximal quantity $b$, then SM$_{i, i+1}$ would fail. Since $\varphi_{1}(p) = 1$ for every $p$, the relevant condition is one on $v$ alone.

\begin{ass}\label{no_ceiling} $\Delta^{+} v(b - 1) < 1$.\footnote{Assumption~\ref{no_ceiling} is not implied by the other conditions. Let $D = \{0, 1, \ldots, 10\}$, $\Theta = \{0.99, 1.99, 2.99\}$ (here, $\gamma = 100$), $v(q) = 50 q - 0.5 q^{2}$, and $p = (\frac{1}{4}, \frac{1}{2}, \frac{1}{4})$, which is strictly log-concave. Then $\varphi_{1}(p) = 1$, $\varphi_{2}(p) = \frac{5}{2}$, and $\varphi_{3}(p) = 6$, while $\Delta^{+} v(q) = 49.5 - q$. No type is shut down, since $\Delta^{+} v(0) = 49.5 > 6 = \varphi_{3}(p)$, and Assumption~\ref{not_too_sdc} holds, since $\Delta^{+} \Delta^{-} v(q) = -1$. Yet $\Delta^{+} v(9) = 40.5$ exceeds all three virtual costs, so that the only solution to~\eqref{combined_FOCs} is $q_i = b = 10$ for every type and strict monotonicity fails. Note that $b = 10 > 6 = \varphi_{3}(p)$ here, so the inequality $b > \varphi_{m}(p)$ on its own does not suffice.}
\end{ass} 

Under Assumption~\ref{no_ceiling} we have $\varphi_{i}(p) \geq i \geq 1 = \varphi_{1}(p) > \Delta^{+} v(b - 1)$ for every type, so that the solutions to~\eqref{combined_FOCs} stay away from the upper end of the grid.

In addition, as previously noted, providing strict incentives for all types means that no type can be shut down---i.e., all types must be served. Notice that this is a joint restriction on $v$ and $p$. Under monotonicity, all types are served if the highest-cost type is served. This observation motivates the following definition.

\begin{defin}[No shutdown]\label{no_shutdown_def} Probability distribution $p \in\Delta(\Theta)$ \textit{shuts down no type} (given $v$) if $\Delta^{+}v(0) > \varphi_{m}(p)$.
\end{defin}

With a continuum of contracts, this condition would be $v'(0) > \varphi_{m}(p)$. In words, the principal's marginal utility at 0 is higher than the virtual cost of the most inefficient type. Thus, awarding said type a positive quantity is profitable. Definition~\ref{no_shutdown_def} is its discrete counterpart, implying that $q_{m}$ must be at least 1.

Notice that the no-shutdown condition is independent of log-concavity. In a scenario with only two types, log-concavity is moot but type $\theta_{2}$ is shut down if $\Delta^{+} v(0) \leq \varphi_{2}(p)$. In general, type $\theta_{i}$ is shut down if $\Delta^{+} v(0) \leq \varphi_{i}(p)$, in which case all types higher than $\theta_{i}$ (if any) are also shut down. With a slight abuse of notation, we denote by $\Delta^{FLN}(\Theta)$ the set of full-support, log-concave distributions that shut down no type (given $v$):
$$\Delta^{FLN}(\Theta):=\left\{p\in \Delta^{FL}(\Theta):\Delta^{+}v(0) > \varphi_{m}(p)\right\}.$$
\begin{rem}\label{nonempty} $\Delta^{FLN}(\Theta) \neq \emptyset$ if and only if $\Delta^{+} v(0) > m$.
\end{rem}

\noindent \textsc{Proof. } See Appendix~\ref{others}. \hfill $\Box$\\

Now, we can ensure strict monotonicity.

\begin{prop}[Strict monotonicity]\label{monotonicity_suf} Under Assumptions~\ref{properties_v_basic} through~\ref{no_ceiling} and for any $p \in \Delta^{FLN}(\Theta)$, the solutions to~\eqref{combined_FOCs} satisfy SM$_{i, i+1}$ for all $i = 1, \ldots, m - 1$. If for some types $i, i + 1$ we have two solutions $q_i, \hat{q}_{i}$ and $q_{i + 1}, \hat{q}_{i+1}$, then $\min\{q_i, \hat{q}_{i}\} > \max\{q_{i + 1}, \hat{q}_{i+1}\}$.
\end{prop}

\noindent \textsc{Proof. } 
Proposition~\ref{monotonicity_weak_suf} establishes weak monotonicity, so it remains to rule out the case $q_{i + 1} = q_i$ for any solutions to the corresponding F.O.C.s. Since $p$ shuts down no type, we can rule out $q_{i + 1} = q_i = 0$, and by Assumption~\ref{no_ceiling} we can rule out $q_{i + 1} = q_i = b$; so both quantities lie in $\{1, \ldots, b - 1\}$. By Assumption~\ref{not_too_sdc},
\begin{eqnarray} 
\Delta^+ \Delta^- v(q_i) & \geq & - 1 \nonumber \\
\Delta^- v(q_i + 1) - \Delta^- v(q_i) & \geq & - 1 \nonumber \\
\Delta^+ v(q_i) - \Delta^- v(q_i) & \geq & - 1 \nonumber \\
\Delta^+ v(q_i) - \Delta^- v(q_i) & > & \left(i+\frac{\sum_{j < i} p_j}{p_i}\right) - \left(i + 1 + \frac{\sum_{j < i + 1} p_j}{p_{i + 1}}\right) \nonumber \\
\Delta^+ v(q_i) - \Delta^- v(q_i) & > & \varphi_{i}(p) - \varphi_{i + 1}(p), \label{Ass2}
\end{eqnarray} 
where the second line restates the first because $\Delta^+ \Delta^- v(q_i) = \Delta^- v(q_i + 1) - \Delta^- v(q_i)$, the third because $\Delta^- v(q_i + 1) = \Delta^+ v(q_i)$, and the strict inequality in the fourth line follows from Corollary~\ref{monotonicity_virtual_valuations}: $\varphi_{i + 1}(p) - \varphi_{i}(p) = 1 + \frac{\sum_{j < i + 1} p_j}{p_{i + 1}} - \frac{\sum_{j < i} p_j}{p_i} > 1$.

Now, suppose that there are unique optimal quantities for each type, and that $q_{i + 1} = q_i$. From~\eqref{combined_FOCs}, $\Delta^{+} v(q_i) \leq \varphi_{i}(p)$ and $\varphi_{i + 1}(p) \leq \Delta^{-} v(q_{i + 1}) = \Delta^{-} v(q_i)$, so that $\Delta^{+} v(q_i) - \Delta^{-} v(q_i) \leq  \varphi_{i}(p) - \varphi_{i + 1}(p)$, a contradiction to inequality~\eqref{Ass2}. Once again, repeat the argument for $\tilde{q}_{i} = \min\{q_i, \hat{q}_{i}\}$ and $\tilde{q}_{i + 1} = \max\{q_{i + 1}, \hat{q}_{i + 1}\}$ in the case of multiplicity.\hfill $\Box$\\ 

Table~\ref{so_far} summarizes and compares the weak and strict formulations of the screening problem, highlighting the relevant constraints and the sufficient conditions for implementability of the solutions to~\eqref{combined_FOCs} in each case.

\begin{table}[h!]
\medskip
\centering
\begin{tabular}{l|l|l}
& Weak inequality constraints & Strict inequality constraints\\ \hline 
Relevant & IC$_{i, j}$: $t_{i} - \theta_{i}q_i \geq t_{j} - \theta_{i}q_{j}$  & SIC$_{i, j}$: $t_{i} - \theta_{i}q_i > t_{j} - \theta_{i}q_{j}$ \\
constraints & PC$_i$: $t_{i} - \theta_{i} q_i \geq 0$ & SPC$_i$: $t_{i} - \theta_{i} q_i > 0$ \\ \hline 
Sufficiency of~\eqref{combined_FOCs} & Assumption~\ref{properties_v_basic} & Assumption~\ref{properties_v_basic}  \\
implied by &  & Assumptions~\ref{not_too_sdc}, \ref{no_ceiling}\\
& $p \in \Delta^{FL}(\Theta)$ & $p \in \Delta^{FLN}(\Theta)$ \\
\end{tabular}
\caption{Weak vs. strict formulations of the screening problem.\label{so_far}}
\medskip
\end{table}

Thus, under Assumption~\ref{properties_v_basic} and $p \in \Delta^{FL}(\Theta)$ for screening with weak inequality constraints, and additionally Assumptions~\ref{not_too_sdc} through~\ref{no_ceiling} and $p \in \Delta^{FLN}(\Theta)$ under strict inequality constraints, we characterize the set of optimal quantities by:
$$Q_i(p) := \{q \in D: \Delta^{+} v(q) \leq \varphi_{i}(p) \leq \Delta^{-} v(q)\}.$$
The conventions $\Delta^{-} v(0) = + \infty$ and $\Delta^{+} v(b) = - \infty$ make one of the two inequalities vacuous at the respective end of the grid: $q = 0$ belongs to $Q_i(p)$ if and only if $\Delta^{+} v(0) \leq \varphi_{i}(p)$, in which case type $\theta_{i}$ is shut down, and $q = b$ belongs to $Q_i(p)$ if and only if $\varphi_{i}(p) \leq \Delta^{-} v(b) = \Delta^{+} v(b - 1)$, in which case the grid of quantities is exhausted. With these conventions $Q_i(p)$ is exactly the set of maximizers of $v(q) - \varphi_{i}(p) q$ over the finite set $D$ (Appendix~\ref{discrete_c_o}) and is therefore nonempty. Recall that sets $Q_i(p)$ are at least singletons and at most doubletons consisting of adjacent quantities. 

Define the set of optimal menus for the discrete screening problems with weak constraints and strict constraints, respectively, by
\begin{align*} \mathcal{M}^*_w(p) & := \left\{\left\{(q_i, t_{i}) : i = 1, \ldots, m\right\} : \begin{array}{l} q_i \in Q_i(p) \mbox{ and } t_{i}\text{ is given by~\eqref{optimal_ts_m}, \eqref{optimal_ts_i}} \\ \mbox{for all } i = 1, \ldots, m, \mbox{ with } \mathbb{I}_s = 0\end{array}\right\}, \\
\mathcal{M}^*_s(p) & := \left\{\left\{(q_i, t_{i}) : i = 1, \ldots, m\right\} : \begin{array}{l} q_i \in Q_i(p) \mbox{ and } t_{i}\text{ is given by~\eqref{optimal_ts_m}, \eqref{optimal_ts_i}} \\ \mbox{for all } i = 1, \ldots, m, \mbox{ with } \mathbb{I}_s = 1\end{array}\right\}.
\end{align*} Note that $\mathcal{M}^*_w(p)$ and $\mathcal{M}^*_s(p)$ differ just in the value of $\mathbb{I}_s$. A menu in either set collects the contracts designed for the $m$ types and therefore contains at most $m$ contracts. Under the assumptions for the strict formulation it contains exactly $m$, since strict monotonicity makes the quantities---and hence the contracts---pairwise distinct. Under the weak formulation, different types may be awarded the same quantity, in which case they are pooled at the same contract: By equation~\eqref{optimal_ts_i} with $\mathbb{I}_s = 0$, $q_i = q_{i + 1}$ implies $t_{i} - t_{i + 1} = \left(i q_i + \sum_{j > i} q_{j}\right) - \left((i + 1) q_{i + 1} + \sum_{j > i + 1} q_{j}\right) = i\left(q_i - q_{i + 1}\right) = 0$. Example~\ref{weak_mono} is a case in point: All three types are awarded $q = 50$ and $t = 150$, so that the optimal menu is the single pooling contract $(50, 150)$. 

For $\gamma \neq \infty$, incentives are strict for free and thus the weak and strict formulations of the screening problem yield the same optimal menus. In this case, we simply write $\mathcal{M}^{*}(p)$ for the set of optimal menus given $p$.

\begin{obs}\label{coincide} Assumptions~\ref{properties_v_basic}, \ref{not_too_sdc} and~\ref{no_ceiling} imply that for all $\gamma \neq \infty$ and every $p \in \Delta^{FLN}(\Theta)$, $\mathcal{M}^*_w(p)  = \mathcal{M}^*_s(p)$.
\end{obs}

The equality is immediate from the two displayed definitions, since $1 - \left\lceil \sfrac{1}{\gamma}\right\rceil = 0$ for $\gamma \neq \infty$; the assumptions are what make the two sets the sets of optimal menus of their respective constrained problems. Without them, $\mathcal{M}^*_s(p)$ as defined above could contain menus that violate strict monotonicity and hence fail strict incentive compatibility, or that award the least efficient type no output, in which case his rounding-up rent $\frac{1}{\gamma}q_{m}$ vanishes and SPC$_{m}$ fails.

We close the section by revisiting multiplicity. We present two main results. The first one (Propositions~\ref{no_indifference} and~\ref{bigger_one}) concerns the agent's choice between contracts, and in particular between the two contracts that arise when a type has two optimal quantities. The second one (Proposition~\ref{unique_contracts}) establishes that uniqueness is generic, and that it is equivalent to robustness with respect to small perturbations of $p$.

We first record a property of our type specification that we will use again in Sections~\ref{sec: equilibrium} and~\ref{sec: rationalizability}: As long as types are off-set from the integers, no type is ever indifferent between two distinct contracts, whether or not they belong to an optimal menu.

\begin{prop}[No indifference]\label{no_indifference} Let $\gamma \neq \infty$. For every $i = 1, \ldots, m$ and any two contracts $\bm{c} = (q, t)$ and $\bm{c}' = (q', t')$ in $D^2$, $u_A(\bm{c}, \theta_{i}) = u_A(\bm{c}', \theta_{i})$ implies $\bm{c} = \bm{c}'$. In particular, for every menu $M \in \mathcal{M}$, $\argmax_{\bm{c} \in M \cup \{\bm{o}\}} u_A(\bm{c}, \theta_{i})$ is a singleton.
\end{prop}

\noindent \textsc{Proof. } Since $\theta_{i} = i - \sfrac{1}{\gamma}$,
$$u_A(\bm{c}, \theta_{i}) - u_A(\bm{c}', \theta_{i}) = \left((t - i q) - (t' - i q')\right) + \frac{q - q'}{\gamma}.$$
The term in parentheses is an integer, while $|q - q'| \leq b < \gamma$ implies $\left|\frac{q - q'}{\gamma}\right| < 1$. Hence the difference vanishes only if $q = q'$, and then only if $t = t'$. Taking $\bm{c}' = \bm{o} = (0, 0)$ shows in addition that the only contract leaving the agent his outside-option value is $\bm{o}$ itself; the singleton claim follows. \hfill $\Box$\\

In particular, when there are two different optimal quantities for the principal, she could offer both; but the agent is not indifferent between the two corresponding contracts. As the rounding-up rent scales with quantity, he strictly prefers the contract with the higher of the two quantities.

\begin{prop}[More is better]\label{bigger_one} Let $\gamma \neq \infty$ and, given $p$, let both $M, \tilde{M} \in \mathcal{M}^*(p)$ be two different optimal menus for the principal for which there exists some type $\theta_{i}$ such that $(q_i, t_{i}) \in M$, $(\tilde{q}_{i}, \tilde{t}_{i})\in \tilde{M}$, $q_i=\tilde{q}_{i} + 1$, and $q_{j} = \tilde{q}_{j}$ for all $j \neq i$. Then, $u_A(q_i, t_{i},\theta_{i}) > u_A(\tilde{q}_{i}, \tilde{t}_{i},\theta_{i})$. 
\end{prop}

\noindent \textsc{Proof. } Substituting the optimal transfers in equations~\eqref{optimal_ts_m} and~\eqref{optimal_ts_i} into $u_A(q_i, t_{i},\theta_{i}) = t_{i} - \theta_{i} q_i$ and using $\theta_{i} = i - \sfrac{1}{\gamma}$, we obtain for every $i = 1, \ldots, m$:
\begin{align*} 
u_A(q_i, t_{i},\theta_{i}) & = i q_i + \sum_{j > i} q_{j} + (m - i + 1) \mathbb{I}_s \left(1 - \left\lceil \frac{1}{\gamma}\right\rceil\right) - \left(i - \frac{1}{\gamma}\right) q_i
\\
& = \frac{1}{\gamma} q_i + \sum_{j > i} q_{j} + (m - i + 1) \mathbb{I}_s \left(1 - \left\lceil \frac{1}{\gamma}\right\rceil\right),
\end{align*} 
where for $i = m$ the sum is zero. Since $q_{j} = \tilde{q}_{j}$ for all $j \neq i$, the sum and the last term coincide across the two menus, so that 
$$u_A(q_i, t_{i},\theta_{i}) - u_A(\tilde{q}_{i}, \tilde{t}_{i},\theta_{i}) = \frac{q_i - \tilde{q}_{i}}{\gamma} = \frac{1}{\gamma} > 0.$$ Note that this is where $\gamma \neq \infty$ is needed: When $\gamma = \infty$, the rent does not depend on $q_i$ at all---for either value of $\mathbb{I}_s$---and the agent is indifferent between the two contracts. \hfill $\Box$\\

Proposition~\ref{nonuniqueness} identifies the necessary and sufficient condition for multiplicity of the optimal quantity for $\theta_{i}$, namely $\Delta^{-}v(\max\{q_{i},q'_{i}\})=\varphi_{i}(p)=\Delta^{+}v(\min\{q_{i},q'_{i}\})$. Clearly, a slight perturbation of $p$ can break either equality and achieve uniqueness. The only exception is type $\theta_{1}$, as his F.O.C.s do not involve the marginal belief. We tackle this special case first, by imposing the following nondegeneracy condition on $v$.

\begin{ass}\label{efficient_unique} For every $q \in \{0, \ldots, b - 1\}$, $\Delta^+ v(q) \neq 1$. 
\end{ass}

For the remaining types, although it is intuitively clear that uniqueness is robust with respect to probability perturbations, we need to take special care to make the statement precise. In the sequel, topological and measure-theoretic statements about subsets of $\Delta(\Theta)$ are understood relative to the affine hull of $\Delta(\Theta)$; that is, relative to the $(m - 1)$-dimensional affine space $\left\{x \in \mathbb{R}^{m} : \sum_{i = 1}^{m} x_i = 1\right\}$. In particular, we say that a property is \textit{generic} if it holds outside a closed set of $(m - 1)$-dimensional Lebesgue measure zero.\footnote{The qualification is not entirely pedantic: As a subset of $\mathbb{R}^{m}$, the simplex $\Delta(\Theta)$ has empty interior and zero measure.} In this sense, having full support is a generic property, as the set of full-support probabilities on $\Theta$ is the interior of $\Delta(\Theta)$ relative to its affine hull.\footnote{We speak of the interior relative to the affine hull, rather than of the \textit{relative interior}, because the latter is a notion from convex analysis and the set of log-concave distributions is \emph{not} convex. With $m = 3$, both $p = (0.6, 0.3, 0.1)$ and $p' = (0.1, 0.3, 0.6)$ are log-concave, while their mixture $\frac{1}{2} p + \frac{1}{2} p' = (0.35, 0.3, 0.35)$ is not, as $0.3 \cdot 0.3 < 0.35 \cdot 0.35$. Intuitively, log-concave sequences are single-peaked, and averaging a decreasing sequence with an increasing one can produce a valley rather than a peak.} 

To state the last result of the section (and for Section~\ref{sec: rationalizability}), we restrict the principal's beliefs further to strict log-concavity:
$$\Delta^{FSLN}(\Theta) := \left\{p \in \Delta^{FLN}(\Theta): p \mbox{ is strictly log-concave}\right\}.$$
This set is open, and it is nonempty under the same condition, since the belief constructed in the proof of Remark~\ref{nonempty} in Appendix~\ref{others} is strictly log-concave. Openness is what the robustness requirement of Section~\ref{sec: rationalizability} calls for, and it is strictness that delivers it: $\Delta^{FLN}(\Theta)$ contains beliefs at which log-concavity holds with equality, and every neighborhood of such a belief contains beliefs that are not log-concave. The strictly log-concave beliefs are nonetheless dense in $\Delta^{FLN}(\Theta)$: The tilt $p^{\varepsilon}_i \propto p_i e^{- \varepsilon i^{2}}$ lowers $\Delta^{+} \Delta^{-} \log p_{i}$ by exactly $2\varepsilon$ at every $i$ and converges to $p$ as $\varepsilon \rightarrow 0$, and since $\varphi_{m}$ is continuous, $p^{\varepsilon}$ shuts down no type for $\varepsilon$ small.\footnote{Here $p^{\varepsilon}_i := p_i e^{-\varepsilon i^{2}} \big/ \sum_{j = 1}^{m} p_j e^{-\varepsilon j^{2}}$, so that $\log p^{\varepsilon}_i = \log p_i - \varepsilon i^{2} - \log \sum_{j} p_j e^{-\varepsilon j^{2}}$. The normalizing term does not depend on $i$ and hence drops out of $\Delta^{+} \Delta^{-}$, while $\Delta^{+} \Delta^{-} i^{2} = (i+1)^{2} - 2 i^{2} + (i-1)^{2} = 2$. Hence $\Delta^{+} \Delta^{-} \log p^{\varepsilon}_{i} = \Delta^{+} \Delta^{-} \log p_{i} - 2 \varepsilon$ for every $i$, and since log-concavity says precisely that these second differences are nonpositive, the tilt turns it into strict log-concavity. The proof of Remark~\ref{nonempty} uses the same device with an additional linear term in the exponent, whose second difference vanishes.}

\begin{prop}\label{unique_contracts} 
\begin{itemize} 
\item[(i)] Under Assumptions~\ref{properties_v_basic} and~\ref{efficient_unique}, $\mathcal{M}^*_w(p)$ is generically a singleton.
\item[(ii)]  Under Assumptions~\ref{properties_v_basic}, \ref{not_too_sdc}, \ref{no_ceiling}, and~\ref{efficient_unique}, for every $p \in \Delta^{FSLN}(\Theta)$, the following statements are equivalent: $(a)$ $\mathcal{M}^*_s(p)$ is a singleton; $(b)$ There exists an open set $O \subseteq \Delta^{FSLN}(\Theta)$ containing $p$ such that $\mathcal{M}^*_s(p') = \mathcal{M}^*_s(p)$ for all $p' \in O$. The same equivalence holds with $\mathcal{M}^*_w$ in place of $\mathcal{M}^*_s$.
\end{itemize} 
\end{prop}

\noindent \textsc{Proof. } Both parts rest on the same observation. Let $p \in \Delta^{FL}(\Theta)$. Since $\mathcal{M}^*_w(p)$ and $\mathcal{M}^*_s(p)$ differ only by shifts in the transfers that do not depend on the quantities, either fails to be a singleton exactly when $Q_i(p)$ is a doubleton for some $i$. This happens exactly when $\varphi_{i}(p) = \Delta^{+} v(q)$ for some $q \in \{0, \ldots, b - 1\}$ (Proposition~\ref{nonuniqueness}). By Assumption~\ref{efficient_unique}, this cannot occur for $i = 1$. For $i \geq 2$, notice the condition $\varphi_{i}(p) = \Delta^{+} v(q)$ is the linear equation:
$$\sum_{j < i} p_j = \left(\Delta^{+} v(q) - i\right) p_i.$$

(i) The set of distributions such that $\mathcal{M}^*_w(p)$ is not a singleton is contained in the following double union:
$$\bigcup_{i = 2}^{m} \ \bigcup_{q = 0}^{b - 1} \left\{p \in \Delta(\Theta) : \sum_{j < i} p_j = \left(\Delta^{+} v(q) - i\right) p_i \right\}.$$
Each of the finitely many members of this union meets $\Delta(\Theta)$ in an affine subspace of dimension at most $m - 2$; hence the union is closed, has empty interior relative to $\Delta(\Theta)$, and is Lebesgue-null in $\Delta(\Theta)$.

(ii) Suppose first that $\mathcal{M}^*_s(p) = \{M\}$ and let $q_i$ be the quantity assigned to $\theta_{i}$ by $M$. Then $Q_i(p) = \{q_i\}$ for every $i$, so that neither $\Delta^{+} v(q_i) = \varphi_{i}(p)$ nor $\Delta^{-} v(q_i) = \varphi_{i}(p)$; together with~\eqref{combined_FOCs} this gives:
$$\Delta^{+} v(q_i) < \varphi_{i}(p) < \Delta^{-} v(q_i) \qquad \mbox{for all } i = 1, \ldots, m.$$
Each $\varphi_{i}(\cdot)$ is continuous on $\text{int}(\Delta(\Theta))$ and $\Delta^{FSLN}(\Theta)$ is relatively open, so these strict inequalities persist on an open set $O \subseteq \Delta^{FSLN}(\Theta)$ containing $p$. For every $p' \in O$ we then have $Q_i(p') = \{q_i\}$ for all $i$, and since transfers are pinned down by the quantities through equations~\eqref{optimal_ts_m} and~\eqref{optimal_ts_i}, $\mathcal{M}^*_s(p') = \{M\}$.

Conversely, suppose $|\mathcal{M}^*_s(p)| > 1$. By the first paragraph there are $i \geq 2$ and $q_i$ with $Q_i(p) = \{q_i, q_i + 1\}$ and $\varphi_{i}(p) = \Delta^{+} v(q_i)$. Any open $O \subseteq \Delta^{FSLN}(\Theta)$ containing $p$ is open relative to the affine hull of $\Delta(\Theta)$ and hence of dimension $m - 1$, so it is not contained in the affine subspace cut out by the linear equation above. It therefore contains some $\tilde{p}$ with $\varphi_{i}(\tilde{p}) \neq \Delta^{+} v(q_i)$ and, by continuity, close enough to $p$ that $Q_i(\tilde{p})$ is a singleton---equal to $\{q_i\}$ if $\varphi_{i}(\tilde{p}) > \Delta^{+} v(q_i)$ and to $\{q_i + 1\}$ otherwise. In either case, $\mathcal{M}^*_s(\tilde{p}) \neq \mathcal{M}^*_s(p)$, so no such $O$ exists. The argument for $\mathcal{M}^*_w$ is identical, so we omit it.\hfill $\Box$\\

We conclude that the optimal menu is unique for generic beliefs, a result that we invoke in Sections~\ref{sec: equilibrium} and~\ref{sec: rationalizability}.

\section{Equilibrium Screening\label{sec: equilibrium}}

In this section, we focus on equilibrium of the discrete screening game. Here, $p$ represents a fixed common prior. Moreover, when type $\theta_i$ is indifferent between the contract $(q_i, t_i)$ designed for him in some $M \in \mathcal{M}^*_w(p)$ and another contract in $M$ or his outside option, it is common knowledge that he follows the equilibrium convention by selecting $(q_i, t_i)$. In this regard, the common prior $p$ allows the agent with type $\theta_i$ to infer which contract in $M \in \mathcal{M}^*_w(p)$ is intended for him; in other words, it fixes the interpretation of the menu, as we illustrate later on.

By Proposition~\ref{no_indifference}, ties between contracts can arise only in the integer case $\gamma = \infty$. This is the case where the equilibrium tie-breaking convention earns its keep, the case that we exclude in our rationalizability analysis of Section~\ref{sec: rationalizability}. For $\gamma<\infty$, the agent's best response to any menu is unique and the tie-breaking convention is moot. (Should several types be awarded the same quantity, the contracts designed for them coincide, so again no tie between \emph{distinct} contracts arises.)

The principal chooses a menu of contracts to offer to the agent. Thus, a strategy for her is simply $s_P = M \in \mathcal{M}$, and her strategy space is $S_P := \mathcal{M}$. The agent's information sets can be identified with tuples $(\theta, M) \in \Theta \times \mathcal{M}$, representing the privately observed type and the menu offered by the principal. His strategies map information sets into either a contract in the offered menu or his outside option; formally, $s_A: \Theta \times \mathcal{M} \longrightarrow \{(q, t) : q, t \in D \}$ with $s_A(\theta, M) \in M \cup \{\bm{o}\}$. Denote by $S_A$ the agent's set of strategies. 

For all $(\theta, M) \in \Theta \times \mathcal{M}$, define the agent's best response correspondence: $$BR_{A}(\theta, M) :=  \argmax_{\bm{c} \in M \cup \{\bm{o}\}} u_{A}(\bm{c},\theta).$$
Given a common prior $p \in \Delta^{FL}(\Theta)$ and under Assumption~\ref{properties_v_basic}, a pair of strategies $(s^*_P, s^*_A)$ is an equilibrium if: 
\begin{itemize} 
\item[(i)] The principal's strategy chooses a menu of contracts that is optimal given $p$ and the weak incentive-compatibility and participation constraints: $s^*_P \in \mathcal{M}^*_w(p)$;
\item[(ii)] The agent's strategy chooses either a best contract or the outside option, resolving indifferences as intended by the principal: For every $i=1,\ldots,m$ and every $M \in \mathcal{M}$,
$$s^*_A(\theta_{i}, M) \in \left\{ \bm{c} \in BR_{A}(\theta_{i}, M) : M \in \mathcal{M}^*_w(p) \mbox{ implies } \bm{c} = (q_i, t_{i})\right\}.$$ 
\end{itemize}  

As mentioned above, the tie-breaking assumption in part (ii) of the definition is stated relative to $\mathcal{M}^*_w(p)$: It tells the agent which contract to select from a menu that is optimal \textit{given} $p$. The common prior is what makes that reference meaningful to both parties, and thereby fixes a common interpretation of the menu the principal offers. That some such anchor is needed is not vacuous, because a given contract can be the contract intended for different types under different beliefs. Example~\ref{2_priors} below exhibits this phenomenon: We have $p \neq p'$ and a contract $(q^0,t^0)$ such that $(q^0,t^0)=(q_i, t_{i}) \in M \in \mathcal{M}_w^*(p)$ and $(q^0,t^0)=(q'_{j}, t'_{j}) \in M' \in \mathcal{M}_w^*(p')$ for some $i \neq j$.\footnote{At least four types are needed for it, which is why the example departs from the three-type setting of Examples~\ref{example_1} and~\ref{example_nonuniqueness}. Since $\varphi_{1}(p) = 1$ for every $p$, the quantity awarded to the most efficient type is the same under every belief, so that $q_1 = q'_1$. Suppose now that $m = 3$ and that $(q_i, t_{i}) = (q'_{j}, t'_{j})$ for some $i \neq j$. If either index is $1$, then by $q_1 = q'_1$ and monotonicity the coincidence forces $q'_1 = \cdots = q'_{j} $ (or $q_1 = \cdots = q_i$), pooling type $1$ with the other type. If $\{i, j\} = \{2, 3\}$, say $i = 2$ and $j = 3$, then $q_2 = q'_3$ and, by equations~\eqref{optimal_ts_m} and~\eqref{optimal_ts_i}, $2 q_2 + q_3 = 3 q'_3$, whence $q_3 = q_2$ and types $2$ and $3$ are pooled under $p$; the case $i = 3$, $j = 2$ is symmetric. In every case the contract is designed for both types under one of the two beliefs, so that the two index sets are not disjoint and the phenomenon does not genuinely arise.}

\begin{ex}\label{2_priors} Let $m = 4$ with integer types, $\Theta = \{1, 2, 3, 4\}$ (that is, $\gamma = \infty$), let $D = \{0, 1, \ldots, 8\}$, and let $v(q) = 8 q - \frac{1}{2} q^{2}$, so that $\Delta^{+} v(q) = \frac{15}{2} - q$ and $\Delta^{-} v(q) = \frac{17}{2} - q$. Assumption~\ref{properties_v_basic} holds, as do Assumptions~\ref{not_too_sdc}, \ref{no_ceiling}, and~\ref{efficient_unique}. Consider the two priors
$$p = \frac{1}{24}\left(1, 4, 13, 6\right) \qquad \mbox{ and } \qquad p' = \frac{1}{24}\left(3, 5, 7, 9\right),$$
both of which have full support and are log-concave, so that $p, p' \in \Delta^{FL}(\Theta)$. Their virtual costs are
$$\varphi(p) = \left(1, \frac{9}{4}, \frac{44}{13}, 7\right) \qquad \mbox{ and } \qquad \varphi(p') = \left(1, \frac{13}{5}, \frac{29}{7}, \frac{17}{3}\right),$$
and~\eqref{combined_FOCs} has a unique solution for every type under either prior:

\medskip
\centerline{$\begin{array}{l|c|c|c|c} & \theta_{1} & \theta_{2} & \theta_{3} & \theta_{4} \\ \hline
\mbox{optimal menu given } p & (7, 19) & (6, 18) & (5, 16) & (1, 4) \\
\mbox{optimal menu given } p' & (7, 18) & (5, 16) & (4, 14) & (2, 8)
\end{array}$}
\medskip

\noindent Both menus award a positive quantity to every type and consist of four distinct contracts. The contract $(5, 16)$ belongs to both. Under $p$ it is the contract designed for $\theta_{3}$, while under $p'$ it is the contract designed for $\theta_{2}$. \hfill $\Box$
\end{ex}

The tie-breaking convention is doing the heavy lifting in the example. Since $\gamma = \infty$, the transfers in equations~\eqref{optimal_ts_m} and~\eqref{optimal_ts_i} with $\mathbb{I}_s = 0$ make $\mbox{IC}_{i, i+1}$ bind with equality, so that every type is indifferent between his own contract and the one designed for the adjacent, less efficient type. Under $p$, type $\theta_{3}$ is indifferent between $(5, 16)$ and $(1, 4)$---both leave him a payoff of $1$---and the convention selects $(5, 16)$; under $p'$, type $\theta_{2}$ is indifferent between $(5, 16)$ and $(4, 14)$---both leave him a payoff of $6$---and the convention again selects $(5, 16)$. The least efficient type is a degenerate instance of the same phenomenon: Since $t_{m} = m q_{m}$ when $\gamma = \infty$, his rent is zero and he is indifferent between his own contract and his outside option; the equilibrium convention favors participating.

Which type a contract is \textit{for} is thus not a property of the contract, but rather of the type distribution that generated the menu containing it. The common prior is what makes that property common knowledge, and hence what the convention in part (ii) of the definition of equilibrium can refer to. 

The theorem below summarizes our analysis of equilibrium screening.

\begin{theo}[Equilibrium screening]\label{unique_equilibrium} Let $v$ satisfy Assumption~\ref{properties_v_basic} and $p \in \Delta^{FL}(\Theta)$ be a common prior. Then there exists an equilibrium of the screening game in which the principal offers a menu $M \in \mathcal{M}_w^*(p)$ and the agent with type $\theta_{i}$ chooses contract $(q_i, t_{i}) \in M$. Moreover, if $v$ also satisfies Assumption~\ref{efficient_unique}, then for almost all $p \in \Delta^{FL}(\Theta)$ the equilibrium menu is unique.
\end{theo}

\noindent \textsc{Proof. } See Propositions~\ref{monotonicity_weak_suf} and~\ref{unique_contracts}(i). \hfill $\Box$\\ 

Note that Theorem~\ref{unique_equilibrium} does not impose any assumptions on $\gamma$. That is, it applies to any $\gamma \in \mathbb{N} \cup \{\infty\}$---and, as observed at the beginning of this section, the tie-breaking convention on which it relies has bite only in the integer case $\gamma = \infty$.

\section{Rationalizable Screening\label{sec: rationalizability}}

In this section, we take a rationalizability approach to screening, which allows us to give up the common prior assumption. Under rationalizability, the principal must form beliefs not only about the agent's type but also about the agent's behavior, since no equilibrium behavior is assumed upfront. In particular, no tie-breaking convention is assumed when the agent is indifferent between two contracts in the menu that is optimal for the marginal belief $p$. Moreover, if---say---type $\theta_i$ selects contract $(q_{i + 1}, t_{i + 1})$, the principal's ``effective'' belief---her belief over which contracts get chosen---shifts some probability mass from the contract designed for $\theta_i$ to the one designed for $\theta_{i+1}$, so that it differs from $p$. Consequently, the original menu would cease to be optimal for $p$. In fact, this effective belief may even violate log-concavity, which would lead to ironing/bunching. 

These complications are avoided once cost types are off-set from the integers. By Proposition~\ref{no_indifference}, no type is then ever indifferent between two distinct contracts, so his choice from any menu is unique and the principal cannot rationalize him choosing the ``wrong'' one. We therefore fix $\gamma<\infty$ in what follows and state the analysis under strict incentive-compatibility and participation constraints; by Observation~\ref{coincide}, the weak and the strict formulations select the same optimal menus for such $\gamma$.
 
The space of strategies is as in Section~\ref{sec: equilibrium}. The agent's beliefs over the principal's strategies are degenerate, as he observes his type and the menu offered by the principal. The principal forms beliefs both about the types and the strategies of the agent; we denote them by $\beta_{P} \in \Delta(\Theta \times S_A)$. The principal's expected payoff from offering menu $M \in \mathcal{M}$ given her belief $\beta_{P} \in \Delta(\Theta \times S_A)$ is defined by: 
$$U_{P}(M,\beta_{P}) :=  \sum_{(\theta, s_A) \in \Theta \times S_A} \beta_{P}(\{(\theta, s_A)\}) \cdot u_{P}\left(s_A(\theta, M)\right).$$ 
Her best response correspondence is given by:
$$BR_{P}(\beta_{P}) := \argmax_{M \in \mathcal{M}} U_{P}(M, \beta_{P}).$$ 

Our rationalizability notion, \textit{$\Delta$-O rationalizability}, combines the idea of $\Delta$-rationalizability---i.e., rationalizability with restrictions on first-order beliefs over types---with robustness of best replies to small perturbations of beliefs. Formally, recall from Section~\ref{sec: screening} the set $\Delta^{FSLN}(\Theta)$ of full-support, strictly log-concave beliefs under which no type is shut down; our set ``$\Delta$'' of beliefs is:
\[
\Delta^{FSLN}(\Theta \times S_A) := \left\{\beta_{P} \in \Delta(\Theta \times S_A): \text{marg}_{\Theta} \ \beta_{P}\in \Delta^{FSLN}(\Theta)\right\}.
\] 
This belief restriction is motivated by our desire to employ the optimal screening contracts of Section~\ref{sec: screening} and facilitate the comparison to ``standard'' equilibrium of screening games derived in the previous section. Requiring robustness of best replies to small perturbations of marginal beliefs over types is what rules out menus that are optimal only by a tie: By Proposition~\ref{unique_contracts}(ii), a menu is optimal throughout an open set of marginal beliefs if and only if it is the uniquely optimal menu for each belief in that set. 

To characterize robustness, we say that a set of beliefs for the principal $O \subseteq \Delta^{FSLN}(\Theta \times S_A)$ is \textit{m-open} if the set of corresponding marginals on $\Theta$ is open (in the usual sense). 

\begin{defin}[m-openness]\label{m_open} Subset $O \subseteq \Delta^{FSLN}(\Theta \times S_A)$ is \textit{m-open} if $\{\text{marg}_\Theta \beta_{P}: \beta_{P} \in O\}$ is an open subset of $\Delta^{FSLN}(\Theta)$.
\end{defin}

We are ready to define our rationalizability procedure. At each level $k$, the game plays out as follows. The principal forms beliefs about the agent's types and strategies such that: (a) Her marginal belief over the agent's types has full support, is strictly log-concave, and precludes shutting down; (b) She is certain of the $(k - 1)$-level $\Delta$-O rationalizable strategies of the agent. She then selects a $(k-1)$-level rationalizable menu which is a best response to all beliefs in some m-open set of $k$-level beliefs. Finally, the agent selects a best contract from the menu offered by the principal as long as his participation constraint is satisfied, and his outside option otherwise. Formally:

\begin{defin}[$\Delta$-O Rationalizability]\label{EFR_full_awareness} Define recursively $R^0_{P} = \mathcal{M}$, $R^0_{A} = S_A$, and for $k \geq 1$, 
\begin{eqnarray*}  
B_P^{k} & = & \left\{\beta_P \in \Delta^{FSLN}(\Theta \times S_A): \text{marg}_{S_A} \ \beta_{P}\left(R_A^{k-1}\right) = 1\right\}; \\
R_{P}^{k} & = & \left\{ M \in R_{P}^{k-1}: \exists \text{ nonempty m-open set } O \subseteq B_P^{k}, \ M \in \bigcap_{\beta_{P} \in O }BR_{P}(\beta_{P})\right\}; \\
R_A^k & = & \left\{ s_A \in R_A^{k - 1}: \forall(\theta, M) \in \Theta \times \mathcal{M}, \  s_A(\theta, M) \in BR_{A}(\theta, M)\right\}.
\end{eqnarray*}
The sets of $\Delta$-O rationalizable strategies of the principal and of the agent are $R_{P}^{\infty} = \bigcap_{k = 1}^{\infty} R_{P}^{k}$ and $R_{A}^{\infty} = \bigcap_{k = 1}^{\infty} R_{A}^{k}$, respectively.
\end{defin}

Even if the optimal menu from Section~\ref{sec: screening} is unique---i.e., if $ \mathcal{M}^*_s(p)$ is a singleton---the principal can always add ``redundant'' contracts, namely contracts that no type of the agent would choose, if it is costless to do so. Thus, to characterize the $\Delta$-O rationalizable menus of contracts offered by the principal, we expand the set of optimal menus $\mathcal{M}^*_s(p)$ as follows: For any $p \in \Delta^{FSLN}(\Theta)$, define:
$$\mathcal{M}_s(p) := \left\{M \in \mathcal{M}:  \begin{array}{l}\exists M'\in \mathcal{M}^*_s(p),\,M' \subseteq M \mbox{ and } \forall \bm{c} \in M \setminus M' \\ \left. \left(\nexists \theta \in \Theta \ (\bm{c} \in \arg \max_{\bm{c}' \in M} u_A(\bm{c}', \theta) \right)\right) \end{array}\right\}.$$ That is, any menu in $\mathcal{M}_s(p)$ includes a menu of optimal contracts in $\mathcal{M}_s^*(p)$ and additional contracts which no type of the agent chooses. 

The next theorem shows that the set of $\Delta$-O rationalizable menus is the set of optimal menus for the corresponding beliefs in $\Delta$, augmented to include irrelevant contracts. Moreover, we ``get there'' quite quickly: The maximal reduction under $\Delta$-O rationalizability is reached after only level-1 reasoning by the agent and level-2 reasoning by the principal.

\begin{theo}[$\Delta$-O Rationalizable menus]\label{benchmark} Fix $\gamma \neq \infty$. Under Assumptions~\ref{properties_v_basic}, \ref{not_too_sdc}, \ref{no_ceiling} and~\ref{efficient_unique}\footnote{Robustness alone yields uniqueness of the optimal contract for types $\theta_{2}, \ldots, \theta_{m}$ only: For $i \geq 2$, multiplicity requires $\varphi_{i}(p) = \Delta^{+} v(q)$, which a perturbation of $p$ breaks. Since $\varphi_{1}(p) = 1$ for every $p$, no perturbation can break a tie at the most efficient type; uniqueness there is what Assumption~\ref{efficient_unique} delivers, just as in Theorem~\ref{unique_equilibrium}.} for $v$, $s^{\infty}_{P} \in R^{\infty}_{P}$ if and only if there exists a nonempty open set $O \subseteq \Delta^{FSLN}(\Theta)$ such that $s^{\infty}_{P} \in \mathcal{M}_s(p)$ for all $p \in O$. Moreover, $R_{A}^{\infty} = R_{A}^{1}$ and $R_{P}^{\infty} = R_{P}^{2}$.
\end{theo}

\noindent \textsc{Proof. } The proof is by induction on levels of elimination.

\noindent {\bf Level 1, principal}. No menu is eliminated at this level: $R_P^1 = \mathcal{M}$. Since $R_A^0 = S_A$, the principal's beliefs about the agent's strategies are entirely unrestricted here. Let $s_{\bm{o}} \in S_A$ be the strategy that selects the outside option at every information set, let $U \subseteq \Delta^{FSLN}(\Theta)$ be nonempty and open, and consider
$$O = \left\{\beta_P \in \Delta^{FSLN}(\Theta \times S_A) : \text{marg}_{S_A} \ \beta_{P} = \delta_{s_{\bm{o}}} \mbox{ and } \text{marg}_{\Theta} \ \beta_{P} \in U\right\}.$$
The set of marginals on $\Theta$ induced by $O$ is $U$, so $O$ is nonempty and m-open; recall that m-openness restricts only the marginals on $\Theta$ (Definition~\ref{m_open}), so beliefs concentrated on a single strategy of the agent are not excluded. For every $\beta_{P} \in O$ and every $M \in \mathcal{M}$ we have $U_{P}(M, \beta_{P}) = u_{P}(\bm{o}) = v(0) - 0 = 0$, whence $BR_{P}(\beta_{P}) = \mathcal{M}$. Thus every menu belongs to $\bigcap_{\beta_{P} \in O} BR_{P}(\beta_{P})$. Nothing is lost by this: only $R_P^2$ enters the characterization below.

\noindent {\bf Level 1, agent}. For any menu offered by the principal, the agent selects the contract that maximizes his payoff given his type among those yielding a nonnegative payoff; if the menu contains no such contract, he selects the outside option. Formally, \\$R_A^1 = \left\{s_A \in S_A : \forall(\theta, M) \in \Theta \times \mathcal{M} \ (s_A(\theta, M) \in BR_{A}(\theta, M))\right\}$.

\noindent {\bf Level 2, principal}. The principal is now certain of $R_A^1$. Thus, for any $\beta_P \in B_P^2$ and any $M \in \mathcal{M}$, she is certain that each type $\theta_{i} \in \Theta$ chooses among $(q_i,t_{i}) \in M\cup\{\bm{o}\}$ such that:
\[
(q_i,t_{i}) \in \argmax_{(q,t) \in M \cup \{\bm{o}\}} \ u_{A}(q,t,\theta_{i}).
\]
If for any $i$ there are multiple optimal choices, the principal can hold any belief about said type randomizing. Then, $U_{P}(M,\beta_{P})$ becomes:
\[
U_{P}(M,\beta_{P})=\sum_{i=1}^{m}\sum_{(q,t)\in M}\, \text{marg}_\Theta \ \beta_{P}\left(\left\{\theta_{i}:(q,t)\in\argmax_{(\tilde{q},\tilde{t})\in M\cup\{\bm{o}\}}\,u_{A}(\tilde{q},\tilde{t},\theta_{i})\right\}\right)u_{P}(q,t).
\]
We claim that $BR_{P}(\beta_{P})=\mathcal{M}_s(\text{marg}_{\Theta}\ \beta_P)$ for every $\beta_{P} \in B_P^2$. Given any menu, the contracts the agent chooses satisfy the weak constraints by revealed preference, and by Observation~\ref{coincide} the weak and the strict problems have the same optimal menus for $\gamma \neq \infty$. Hence $M\in BR_{P}(\beta_{P})$ for every $M\in\mathcal{M}^{*}_s(\text{marg}_{\Theta}\ \beta_P)$, and hence for every $M\in\mathcal{M}_s(\text{marg}_{\Theta}\ \beta_P)$: Adding to a menu contracts that no type of the agent chooses leaves $U_{P}(\cdot, \beta_{P})$ unchanged. Conversely, take $M\in BR_{P}(\beta_{P})$. For each $i=1,\ldots,m$, by Proposition~\ref{no_indifference}, $\argmax_{(q,t) \in M \cup \{\bm{o}\}}\,u_{A}(q,t,\theta_{i})$ is a singleton even for types whose F.O.C.s have two solutions. Denote by $(q_i,t_{i})$ the unique element of this set for each $i=1,\ldots,m$---notice that, with $\beta_{P} \in \Delta^{FSLN}(\Theta \times S_A)$, no type chooses the outside option. Strict monotonicity in the quantities is ensured for $\beta_{P} \in \Delta^{FSLN}(\Theta \times S_A)$ and Assumptions~\ref{properties_v_basic}, \ref{not_too_sdc}, and~\ref{no_ceiling} by Proposition~\ref{monotonicity_suf}. Thus, $\{(q_i,t_{i}):i=1,\ldots,m\}\in\mathcal{M}^{*}_s(\text{marg}_{\Theta}\ \beta_P)$. Now, take any $(q,t)\in M\setminus \{(q_i,t_{i}):i=1,\ldots,m\}$; since $(q_i,t_{i})$ is the unique best response of type $\theta_{i}$ in $M\cup\{\bm{o}\}$ for each $i=1,\ldots,m$, there is no type for whom $(q,t)$ is a best response. Thus, $M\in\mathcal{M}_s(\text{marg}_{\Theta}\ \beta_P)$.

Being a best response to a single belief is not enough, however: by Definition~\ref{EFR_full_awareness}, a menu must be a best response to all beliefs in an m-open set. We conclude that:
$$R_P^2 = \bigcup_{\left\{p \, \in \, \Delta^{FSLN}(\Theta) \; : \; \left|\mathcal{M}^{*}_s(p)\right| \, = \, 1\right\}} \mathcal{M}_s(p).$$
To see this, note first that which contracts of a menu no type of the agent chooses does not depend on $p$, so that $\mathcal{M}_s(p)$ is determined by $\mathcal{M}^{*}_s(p)$. If $\mathcal{M}^{*}_s(p)$ is a singleton, then by Proposition~\ref{unique_contracts}(ii) there is an open set $O \subseteq \Delta^{FSLN}(\Theta)$ containing $p$ on which $\mathcal{M}^{*}_s$---and hence $\mathcal{M}_s$---is constant; the set of beliefs with marginal on $\Theta$ in $O$ is m-open, so every $M \in \mathcal{M}_s(p)$ belongs to $R_P^2$. Conversely, if $M \in R_P^2$, then $M \in \mathcal{M}_s(p')$ for all $p'$ in some open $O \subseteq \Delta^{FSLN}(\Theta)$; by Observation~\ref{coincide} and Proposition~\ref{unique_contracts}(i) the beliefs at which the optimal menu is not unique are nowhere dense, so $O$ contains some $p$ with $\left|\mathcal{M}^{*}_s(p)\right| = 1$, and $M \in \mathcal{M}_s(p)$.

\noindent {\bf Level 2, agent}. No further change. No matter what menu the agent is presented with, he chooses a best response and thus $R_A^2 = R_A^1$. 

\noindent {\bf Level 3, principal}. No further change. Since $R_A^2 = R_A^1$, we must have $B_P^3 = B_P^2$. Thus, $R_P^3 = R_P^2$.  

\noindent {\bf Level 3, agent}. No further change. No matter what menu the agent is presented with, he chooses a best response and thus $R_A^3 = R_A^2$.  

The maximal reduction of strategies is reached after two levels of the procedure; thus, we have: $R_{A}^{\infty} = R_{A}^{2} = R_{A}^{1}$, $R_{P}^{\infty} = R_{P}^{2}$. This concludes the proof. \hfill $\Box$\\

Note that the agent's side of the procedure collapses entirely. Since Theorem~\ref{benchmark} fixes $\gamma \neq \infty$, Proposition~\ref{no_indifference} makes $BR_{A}(\theta, M)$ a singleton at every information set $(\theta, M) \in \Theta \times \mathcal{M}$, so that $R_{A}^{\infty} = R_{A}^{1}$ is not merely reached after one round but consists of a \emph{single} strategy: From every menu, select the unique strictly preferred contract or the outside option, whichever is (strictly) better. This same strategy is also the agent's equilibrium strategy, because part (ii) of the definition of equilibrium likewise requires a best response at every information set. The tie-breaking convention is moot under $\gamma<\infty$, whether we solve the game by equilibrium or by $\Delta$-O rationalizability.


\section{Discussion and Conclusion}\label{sec: discussion}
 
We begin by comparing the two solution concepts. Under the assumptions of Theorem~\ref{benchmark}, the $\Delta$-O rationalizable outcomes and the equilibrium outcomes coincide up to contracts that no type of the agent would ever choose.

\begin{cor} Under the assumptions of Theorem~\ref{benchmark}, for every $\Delta$-O rationalizable outcome $(s_P, s_A) \in R_P^{\infty} \times R_A^{\infty}$, there exists a nonempty open set $O \subseteq \Delta^{FSLN}(\Theta)$ such that $(s_P^*, s_A^*)$ is the unique equilibrium for all $p \in O$ and $s_P^* \subseteq s_P$, $s_A(\theta, s_P) = s^*_A(\theta, s_P^*)$. Conversely, for every nonempty open set $O \subseteq \Delta^{FSLN}(\Theta)$ and every strategy profile $(s_P^*, s_A^*)$ that is an equilibrium for every $p \in O$, \emph{every} menu $s_P \in \mathcal{M}$ with $s_P^* \subseteq s_P$ such that no contract in $s_P \setminus s_P^*$ maximizes $u_A(\cdot, \theta)$ over $s_P$ for any $\theta \in \Theta$ satisfies $(s_P, s_A) \in R_P^{\infty} \times R_A^{\infty}$ and $s_A(\theta, s_P) = s^*_A(\theta, s_P^*)$, where $s_A$ is the unique element of $R_A^{\infty}$. In particular, $s_P = s_P^*$ qualifies.
\end{cor}

\noindent \textsc{Proof. } Recall from the discussion following Theorem~\ref{benchmark} that $R_A^{\infty}$ consists of a single strategy, which is also the agent's equilibrium strategy; we write $s^*_A$ for it. Moreover, $\Delta^{FSLN}(\Theta) \subseteq \Delta^{FL}(\Theta)$ and every $p \in \Delta^{FSLN}(\Theta)$ shuts down no type, so that Observation~\ref{coincide} applies and $\mathcal{M}^*_w(p) = \mathcal{M}^*_s(p) = \mathcal{M}^{*}(p)$ for every such $p$.

\noindent $(\Rightarrow)$ Let $(s_P, s_A) \in R_P^{\infty} \times R_A^{\infty}$, so that $s_A = s^*_A$. By Theorem~\ref{benchmark}, there is a nonempty open set $O_{0} \subseteq \Delta^{FSLN}(\Theta)$ with $s_P \in \mathcal{M}_s(p)$ for all $p \in O_{0}$. By Proposition~\ref{unique_contracts}(i), the beliefs at which $\mathcal{M}^{*}(p)$ is not a singleton form a closed set with empty interior, so we may pick $p_{0} \in O_{0}$ with $\mathcal{M}^{*}(p_{0}) = \{M^{*}\}$; by Proposition~\ref{unique_contracts}(ii), there is then an open set $O \subseteq O_{0}$ containing $p_{0}$ on which $\mathcal{M}^{*}$ is constantly equal to $\{M^{*}\}$. Set $s_P^{*} := M^{*}$. Since $s_P \in \mathcal{M}_s(p_{0})$ and $\mathcal{M}^{*}_s(p_{0}) = \{M^{*}\}$, we have $s_P^{*} \subseteq s_P$. For every $p \in O$, the profile $(s_P^{*}, s^*_A)$ is an equilibrium by Theorem~\ref{unique_equilibrium}, and it is the unique one: The menu is unique because $\mathcal{M}^*_w(p) = \{M^{*}\}$, and the agent's strategy is unique by the first paragraph. Finally, by the definition of $\mathcal{M}_s(p_{0})$, no contract in $s_P \setminus M^{*}$ maximizes $u_A(\cdot, \theta)$ over $s_P$, whence $BR_{A}(\theta, s_P) = BR_{A}(\theta, M^{*})$ and therefore $s_A(\theta, s_P) = s^*_A(\theta, s_P^{*})$.

\noindent $(\Leftarrow)$ Let $O \subseteq \Delta^{FSLN}(\Theta)$ be nonempty and open, let $(s_P^{*}, s_A^{*})$ be an equilibrium for every $p \in O$, and let $s_P$ be as in the statement. Then $s_P^{*} \in \mathcal{M}^*_w(p) = \mathcal{M}^{*}_s(p)$ for every $p \in O$. Whether a contract of a given menu maximizes $u_A(\cdot, \theta)$ over that menu for some $\theta$ does not depend on $p$, so $s_P \in \mathcal{M}_s(p)$ for every $p \in O$, with $s_P^{*}$ serving as the required element of $\mathcal{M}^{*}_s(p)$; hence $s_P \in R_P^{\infty}$ by Theorem~\ref{benchmark}. Since no contract in $s_P \setminus s_P^{*}$ maximizes $u_A(\cdot, \theta)$ over $s_P$, we again have $BR_{A}(\theta, s_P) = BR_{A}(\theta, s_P^{*})$ and therefore $s_A(\theta, s_P) = s^*_A(\theta, s_P^{*})$ for the unique $s_A \in R_A^{\infty}$, namely $s_A = s^*_A$. Taking $s_P = s_P^{*}$ shows that the class of menus in question is nonempty. \hfill $\Box$\\

Thus, under $\gamma \neq \infty$ and for generic $p$, the set of $\Delta$-O rationalizable menus is the set of equilibrium menus, modulo redundant contracts. We obtain essentially the same result from $\Delta$-O rationalizability at the ``cost'' of the additional Assumptions~\ref{not_too_sdc} and~\ref{no_ceiling}, restricting beliefs to $\Delta^{FSLN}(\Theta)$, and ruling out integer types, but with the benefit of giving up the common prior assumption and avoiding the equilibrium tie-breaking convention. 

Table~\ref{assumptions_table} summarizes which assumptions and conditions each of our two approaches requires. The requirement $b > m$ is part of the model, so that the principal can offer as many as $m$ distinct contracts. No result of the equilibrium analysis relies on it, since weak monotonicity permits pooling; under strict inequality constraints it need not be assumed at all, being implied by Assumptions~\ref{not_too_sdc} and~\ref{no_ceiling} together with the no-shutdown condition.\footnote{Since no type is shut down, $\Delta^{+} v(0) > \varphi_{m}(p)$, while Assumption~\ref{no_ceiling} gives $\Delta^{+} v(b - 1) < 1$. Hence $\varphi_{m}(p) - 1 < \Delta^{+} v(0) - \Delta^{+} v(b - 1) = \sum_{k = 1}^{b - 1} \left(\Delta^{+} v(k - 1) - \Delta^{+} v(k)\right) \leq b - 1$, where the last inequality holds because Assumption~\ref{not_too_sdc} says that $\Delta^{+} v$ falls by at most one unit per step. Thus $b > \varphi_{m}(p) \geq m$, which is the sharp form of our informal requirement that $b$ be large compared to $m$.}

\begin{table}[h!]
\medskip
\centering
\begin{tabular}{l|c|c}
 & Equilibrium & Rationalizability \\
 & (weak inequality & (strict inequality \\
 & constraints) & constraints) \\ \hline
Assumption~\ref{properties_v_basic} on $v$ & Yes & Yes \\
Assumption~\ref{not_too_sdc}: $\Delta^{+} \Delta^{-} v(q) \geq -1$ & No & Yes \\
Assumption~\ref{no_ceiling}: $\Delta^{+} v(b - 1) < 1$ & No & Yes \\
Assumption~\ref{efficient_unique}: $\Delta^{+} v(q) \neq 1$ & Only for uniqueness & Yes \\
Conventions $\Delta^{-} v(0) = + \infty$, $\Delta^{+} v(b) = - \infty$ & Yes & Yes \\ \hline
Full support, $p \in \text{int}(\Delta(\Theta))$ & Yes & Yes \\
Log-concavity, $p \in \Delta^{FL}(\Theta)$ & Yes & Yes \\
Strict log-concavity, $p \in \Delta^{FSLN}(\Theta)$ & No & Yes \\
No shutdown, $\Delta^{+} v(0) > \varphi_{m}(p)$ & No & Yes \\
$\Delta^{+} v(0) > m$ (Remark~\ref{nonempty}) & No & Yes \\ \hline
Non-integer types, $\gamma \neq \infty$ & No & Yes \\ \hline
$\gamma > b$ & Yes & Yes \\
$b > m$ & Assumed, not required & Implied \\ \hline
Common prior & Yes & No \\
Equilibrium tie-breaking convention & Yes & No \\
Robustness on m-open sets of beliefs & No & Yes \\
\end{tabular}
\caption{Equilibrium vs. rationalizability approaches: assumptions and conditions.\label{assumptions_table}}
\medskip
\end{table}
 
Throughout this paper, we discuss robustness with respect to beliefs but we focus on a linear cost structure. What can we say under a more general cost structure? While many results extend and the core of the rationalizability analysis remains the same, the characterization of optimal contracts can change drastically. We elaborate on this point in Appendix~\ref{non_linear_app}; screening becomes a finite-horizon, discrete dynamic programming problem, and we provide an example where monotonicity of the quantities fails even under strictly increasing virtual costs and one where there is an \textit{upward} distortion at the top---which cannot happen in the linear-cost benchmark.
 
We close with an observation about rationalizability and weak incentives. In equilibrium analysis, we can always invoke the tie-breaking convention. But what is the set of rationalizable contracts when we only impose weak incentives? Consider integer types, $\gamma = \infty$. The optimal transfers then leave each type indifferent between his own contract and the one designed for the adjacent, less efficient type, and they leave the highest-cost type indifferent between his contract and his outside option. Hence rationality alone cannot rule out (local) over-reporting, nor rejection by the highest-cost type---which is akin to resolving indifference in the way that most favors the agent rather than the principal. Accordingly, the principal can assign positive probability to strategies of the agent where he chooses ``the wrong'' contract, which can impact the optimal menu. 

\appendix 

\appendixnumbering


\section{Discrete Concave Optimization}\label{discrete_c_o}

We briefly review the basic elements of discrete concave optimization employed in the paper. Denote by $\mathbb{N} = \{1, 2, 3, \ldots\}$ the set of natural numbers, and set $\mathbb{N}_0 := \{0\} \cup \mathbb{N}$. A function $f: \mathbb{N}_0 \longrightarrow \mathbb{R}$ is discrete concave if for all $x \in \mathbb{N}$, $$f(x + 1) + f(x - 1) \leq 2 f(x).$$ It is discrete strictly concave if the inequality holds strictly.

Although we state the results of this appendix for functions on $\mathbb{N}_0$, they apply verbatim to a function defined on a finite grid $D = \{0, 1, \ldots, b\}$, as in the body of the paper, once maximizers are taken over $D$ and the conventions $\Delta^{-} f(0) := + \infty$ and $\Delta^{+} f(b) := - \infty$ are adopted: These conventions make the respective ``first-order condition'' vacuous at each end of the grid, and every chain of inequalities in the proofs below simply terminates at $0$ or at $b$ instead of continuing indefinitely. In particular, on a finite grid a local maximizer of a discrete concave $f$ is a global maximizer of $f$ on that grid, and the set of maximizers is nonempty.

We say that $x$ is a local maximizer of $f$ if $f(x) \geq \max\{f(x + 1), f(x - 1)\}$. It is a strict local maximizer if the inequality holds strictly. We say that $x$ is a global maximizer of $f$ if $f(x) \geq f(y)$ for all $y \in \mathbb{N}_0$. It is a unique global maximizer if the inequality holds strictly for all $y \neq x$. 

\begin{prop}\label{local_vs_global} If $f$ is discrete concave, then $x$ is a global maximizer if and only if it is a local maximizer. Moreover, $x$ is the unique global maximizer if and only if it is a strict local maximizer. 
\end{prop}

\noindent \textsc{Proof. } Let $f$ be discrete concave. The ``only if'' direction is trivial. We prove the ``if'' direction. For any $n\in\mathbb{N}$, $$0 \leq f(x) - f(x + 1) \leq f(x + 1) - f(x + 2) \leq f(x + 2) - f(x + 3) \leq \cdots \leq f(x + n - 1) - f(x + n),$$ where the first inequality follows from $x$ being a local maximizer of $f$ and the other inequalities follow from discrete concavity. This implies $f(x) \geq f(x + 1)$, $f(x+1) \geq f(x + 2)$, $f(x + 2) \geq f(x + 3)$, \ldots, $f(x + n - 1) \geq f(x + n)$. We conclude $f(x) \geq f(x + n)$ for any $n\in\mathbb{N}$. 

Similarly, for any $n\in\mathbb{N}$, $$0 \geq f(x - 1) - f(x) \geq f(x - 2) - f(x - 1) \geq f(x - 3) - f(x - 2) \geq \cdots \geq f(x - n + 1) - f(x - n),$$ where the first inequality follows from $x$ being a local maximizer of $f$ and other inequalities follow from discrete concavity. This implies $f(x) \geq f(x - 1)$, $f(x - 1) \geq f(x - 2)$, $f(x - 2) \geq f(x - 3)$, \ldots, $f(x - n + 1) \geq f(x - n)$. We conclude $f(x) \geq f(x - n)$ for any $n\in\mathbb{N}$. Thus $x$ is a global maximizer. 

The version of the argument for the unique global maximizer follows from strict local maximizer implying that the first inequality in each chain above is a strict inequality. \hfill $\Box$ \\

Define the discrete forward derivative of $f$ at $x \in \mathbb{N}_0$ by: $$\Delta^+ f(x) = f(x + 1) - f(x),$$ and the discrete backward derivative of $f$ at $x \in \mathbb{N}$ by: $$\Delta^- f(x) = f(x) - f(x - 1).$$ Clearly, for any $x \in \mathbb{N}_0$, $$\Delta^+ f(x) = \Delta^- f(x+1),$$ and for any $x \in \mathbb{N}$, $$\Delta^- f(x) = \Delta^+ f(x - 1).$$ 

The next observation is immediate from the definitions.

\begin{obs}\label{d_monot} The following statements are equivalent: 
\begin{enumerate}
\item Function $f$ is monotone nondecreasing, i.e., $x \geq y$ implies $f(x) \geq f(y)$.
\item $\Delta^+ f(x) \geq 0$ for all $x \in \mathbb{N}_0$.
\item $\Delta^- f(x) \geq 0$ for all $x \in \mathbb{N}$. 
\end{enumerate} It is monotone strictly increasing if and only if the inequalities hold strictly. 
\end{obs}

The following proposition on ``first-order conditions'' for optimization involving discrete derivatives is also immediate from the definitions.

\begin{prop}[``First-order conditions''] We have $x \in \mathbb{N}_0$ is a local maximizer of $f$ if and only if: $$\Delta^- f(x) \geq 0 \mbox{ and } \Delta^+ f(x) \leq 0$$ whenever these expressions are defined for $x$. It is a strict local maximizer if and only if the inequalities hold strictly whenever these expressions are defined for $x$. 
\end{prop}

Denote by $\Delta^+ \Delta^+ f(x)$ the second discrete forward derivative of $f$ at $x$ defined by $\Delta^+ (\Delta^+ f(x))$ (and analogously for $\Delta^- \Delta^- f(x)$, $\Delta^- \Delta^+ f(x)$, and $\Delta^+ \Delta^- f(x)$).

\begin{claim} For any $x \in \mathbb{N}$, $\Delta^+ \Delta^- f(x) = \Delta^- \Delta^+ f(x)$. 
\end{claim}

\noindent \textsc{Proof.} $\Delta^+ \Delta^- f(x) = \Delta^-f(x + 1) - \Delta^-f(x) = f(x + 1) - f(x) - f(x) + f(x - 1) = \Delta^+ f(x) - \Delta^+ f(x - 1) = \Delta^- \Delta^+ f(x)$. \hfill $\Box$\\

\begin{obs}\label{second_derivative} The following statements are equivalent: 
\begin{enumerate} 
\item The function $f$ is discrete concave.
\item For all $x \in \mathbb{N}$, $\Delta^+ f(x) \leq \Delta^- f(x).$
\item For all $x \in \mathbb{N}_0$, $\Delta^+ \Delta^+ f(x) \leq 0.$
\item For all $x \in \mathbb{N}$, $\Delta^- \Delta^- f(x + 1) \leq 0.$
\item For all $x \in \mathbb{N}$, $\Delta^+ \Delta^- f(x) \leq 0.$
\item For all $x \in \mathbb{N}$, $\Delta^- \Delta^+ f(x) \leq 0.$
\end{enumerate} The inequalities hold strictly if and only if $f$ is discrete strictly concave.
\end{obs} 

\begin{lem}\label{monotoncity_derivative} If $f$ is discrete concave, then $x > y$ implies: 
\begin{itemize}
\item[(i)] $\Delta^- f(x) \leq \Delta^- f(y)$;
\item[(ii)] $\Delta^+ f(x) \leq \Delta^+ f(y)$;
\item[(iii)] $\Delta^+ f(x) \leq \Delta^- f(y)$;
\item[(iv)] $\Delta^- f(x) \leq \Delta^+ f(y)$.   
\end{itemize} 
\end{lem}  

\noindent \textsc{Proof. } (i) W.l.o.g., let $y \in \mathbb{N}$ and take $x := y + n$ for some $n \in \mathbb{N}$. Observe that: $$f(x) - f(x - 1) \leq f(x - 1) - f(x - 2) \leq \cdots \leq f(y) - f(y - 1),$$ where each inequality follows from $f$ being discrete concave. This chain of inequalities is equivalent to: $$\Delta^- f(x) \leq \Delta^- f(x - 1) \leq \cdots \leq \Delta^- f(y).$$ We conclude (i). 

Item (ii) follows analogously. 

For (iii), apply (2.) from Observation~\ref{second_derivative} to either (i) or (ii). 

For (iv), note that by the definitions of discrete derivatives, $\Delta^- f(x) = \Delta^+ f(x - 1)$. The conclusion follows now from (ii).\hfill $\Box$\\

Note that for $y = x - 1$, property (iv) holds with equality irrespective of whether $f$ is discrete concave.  

\begin{prop}\label{monotoncity_derivative_strict} Let $f$ be discrete strictly concave. 
\begin{itemize}
\item[(i)] $x > y$ if and only if $\Delta^- f(x) < \Delta^- f(y)$;
\item[(ii)] $x > y$ if and only if $\Delta^+ f(x) < \Delta^+ f(y)$;
\item[(iii)] $x > y$ if and only if $\Delta^+ f(x) < \Delta^- f(y)$ and $x \neq y$;
\item[(iv)] $x > y + 1$ if and only if $\Delta^- f(x) < \Delta^+ f(y)$.   
\end{itemize} 
\end{prop} 

\noindent \textsc{Proof. } The ``only if'' conditions for (i) and (ii) follow by analogous arguments as in the proof of Lemma~\ref{monotoncity_derivative}, replacing discrete concavity with discrete strict concavity. 

(i) If: Suppose to the contrary that $\Delta^- f(x) < \Delta^- f(y)$ and $x \leq y$. If $x < y$, then $\Delta^- f(x) > \Delta^- f(y)$ by the ``only if'' condition. If $x = y$, then $\Delta^- f(x) = \Delta^- f(y)$. In either case, we obtain a contradiction. 

(ii) If: Follows analogously. 

(iii) If: Suppose to the contrary that $\Delta^+ f(x) < \Delta^- f(y)$ and $x < y$. Then by Lemma~\ref{monotoncity_derivative}(iv), $\Delta^+ f(x) \geq \Delta^- f(y)$, a contradiction. 

Only if: If $x > y$, apply (2.) from Observation~\ref{second_derivative} to either (i) or (ii).

(iv) If: Suppose to the contrary that $\Delta^- f(x) < \Delta^+ f(y)$ and $x \leq y + 1$. If $x = y + 1$, then $\Delta^- f(x) = \Delta^+ f(y)$ 
by definition. If $x < y + 1$, then $\Delta^- f(x) > \Delta^+ f(y)$ by (iii). In both cases, we have a contradiction. 

Only if: Analogously to the proof of (iv) of Lemma~\ref{monotoncity_derivative} using discrete strict concavity instead of discrete concavity. \hfill $\Box$

\begin{cor} If $f$ is discrete concave, then $x$ satisfying ``first-order conditions'' is sufficient for $x$ being a global maximizer. 
\end{cor}

Discrete strict concavity is \emph{not} sufficient for the global maximizer to be unique. A simple counterexample is $f(x) = - \left(x - \frac{5}{2}\right)^{2}$: we have $\Delta^{+} \Delta^{-} f(x) = -2$ for every $x \in \mathbb{N}$, so $f$ is discrete strictly concave, yet the set of maximizers is $\{2, 3\}$. However, the argmax cannot consist of more than two adjacent elements, as Proposition~\ref{nonuniqueness_gen} establishes.

\begin{prop}\label{nonuniqueness_gen} If $f$ is discrete strictly concave, then there are at most two global maximizers. If $x$ and $y$ are global maximizers, then either $x = y$ or $\max\{x, y\} = \min\{x, y\} + 1$. Moreover, if $x > y$, then $x$ and $y$ are both global maximizers if and only if $\Delta^- f(x) = 0 = \Delta^+ f(y)$. 
\end{prop}

\noindent \textsc{Proof. } Suppose there exist $x, y \in \mathbb{N}_0$ such that first-order conditions are satisfied:
\begin{eqnarray*} \Delta^+ f(x) \leq & 0 & \leq \Delta^{-} f(x), \\
\Delta^+ f(y) \leq & 0 & \leq \Delta^{-} f(y).
\end{eqnarray*} By the previous corollary, both $x$ and $y$ must be global maximizers since $f$ is discrete concave. It follows that: 
\begin{eqnarray*} \Delta^+ f(x) \leq & 0 & \leq \Delta^{-} f(y), \\
\Delta^+ f(y) \leq & 0 & \leq \Delta^{-} f(x).
\end{eqnarray*} W.l.o.g., assume $x > y$. If $x > y + 1$, then by Proposition~\ref{monotoncity_derivative_strict}(iv), $\Delta^- f(x) < \Delta^{+} f(y)$, a contradiction to the last inequality. If $x = y + 1$, then $\Delta^- f(x) = \Delta^- f(y + 1) = \Delta^{+} f(y)$ by the definition of discrete derivatives. It follows that $\Delta^- f(x) = 0 = \Delta^{+} f(y)$. 

Conversely, assume $x > y$ and $\Delta^- f(x) = 0 = \Delta^{+} f(y)$. Since $f$ is discrete strictly concave, apply (2.) from Observation~\ref{second_derivative} at $x$ and at $y$ to get $\Delta^+ f(x) < \Delta^- f(x) = 0$ and $\Delta^- f(y) > \Delta^+ f(y) = 0$. Thus, the first-order conditions must hold.\hfill $\Box$\\


Finally, we will require as an ingredient to our discrete analysis the floor and ceiling functions defined for $x \in \mathbb{R}$ by:
\begin{eqnarray*} \lfloor x \rfloor & := & \max \{n \in \mathbb{Z}: n \leq x \}, \\
\lceil x \rceil & := & \min \{n \in \mathbb{Z} : x \leq n \}.
\end{eqnarray*}

We collect without proof well-known properties of the floor and ceiling functions:

\begin{prop}\label{ceiling} For any $x, y \in \mathbb{R}$ and $n \in \mathbb{Z}$, 
\begin{itemize}
\item[(i)] $x - 1 \leq \lfloor x \rfloor \leq x \leq \lceil x \rceil \leq x + 1$;
\item[(ii)] $\lceil x \rceil = - \lfloor - x \rfloor$; 
\item[(iii)] $\lceil x \rceil + \lceil y \rceil - 1 \leq \lceil x + y \rceil \leq \lceil x \rceil + \lceil y \rceil$;
\item[(iv)] $\lfloor x \rfloor + \lfloor y \rfloor \leq \lfloor x + y \rfloor \leq \lfloor x \rfloor + \lfloor y \rfloor  + 1$;
\item[(v)] $\lfloor x + n \rfloor  = \lfloor x \rfloor + n$;
\item[(vi)] $\lceil x + n \rceil = \lceil x \rceil + n$.  
\end{itemize}
\end{prop}

\appendixnumbering


\section{Constraint Simplifications}\label{const_simp}

In this appendix, we show how several of the standard constraint-simplification results can be extended to our discrete setting. The arguments are virtually identical for both the weak and strict cases, so we focus on the latter. We start with the usual monotonicity and the reduction of participation constraints.

\begin{prop}\label{monot_usual} Under the SIC constraints, $q_i>q_{i + 1}$ for all $i=1,\ldots, m-1$.
\end{prop}

\noindent \textsc{Proof.} Take any $i=1,\ldots, m-1$. By SIC$_{i,i+1}$ and SIC$_{i+1,i}$, 
\[
\theta_{i} q_{i + 1} - \theta_{i} q_i> t_{i + 1} - t_{i}>\theta_{i+1} q_{i + 1} - \theta_{i+1} q_i. 
\]
Thus, $(\theta_{i}- \theta_{i+1})(q_{i + 1} - q_i)>0$, which establishes the proposition given that $\theta_{i}<\theta_{i+1}$.\hfill $\Box$

\begin{prop}\label{IR} SPC$_m$ and the local constraints SIC$_{k, k+1}$, $k = 1, \ldots, m-1$, imply SPC$_i$ for all $i = 1, \ldots, m-1$.
\end{prop}

\noindent \textsc{Proof.} We establish SPC$_i$ for all $i = 1, \ldots, m-1$ by induction on $i$, downwards from $i = m - 1$.

\noindent Base case $i = m - 1$: By SPC$_m$,
\begin{align*}
0 < t_{m} - \theta_{m} q_{m}&\leq t_{m} - \theta_{m-1} q_{m}\hspace{4mm}\text{since }\theta_{m}>\theta_{m-1}\text{ and }q_{m}\geq 0
\\
&< t_{m-1} - \theta_{m-1} q_{m-1}\hspace{4mm}\text{by SIC$_{m-1,m}$},
\end{align*}
which is SPC$_{m-1}$.

\noindent Induction hypothesis: For some $i$ with $1 \leq i \leq m - 2$, SPC$_{i+1}$ holds, that is, $t_{i + 1} - \theta_{i+1} q_{i + 1} > 0$.

\noindent Induction step: We show that SPC$_{i}$ holds. Indeed,
\begin{align*}
0 < t_{i + 1} - \theta_{i+1} q_{i + 1}&\leq t_{i + 1} - \theta_{i} q_{i + 1}\hspace{4mm}\text{since }\theta_{i+1}>\theta_{i}\text{ and }q_{i + 1}\geq 0
\\
&< t_{i} - \theta_{i} q_i\hspace{4mm}\text{by SIC$_{i,i+1}$},
\end{align*}
which is SPC$_{i}$. Downward induction from $i = m - 1$ to $i = 1$ therefore yields SPC$_i$ for all $i = 1, \ldots, m - 1$.\hfill $\Box$\\



The next propositions establish that we can restrict attention to the ``local'' SIC constraints, SIC$_{i, i+1}$ and SIC$_{i, i-1}$. The two inductions run in opposite directions, so we prove both.

\begin{prop}\label{USIC} Assume that $q_i>q_{i + 1}$ for all $i =1,\ldots,m-1$. Then the local constraints SIC$_{k, k+1}$, $k = 1, \ldots, m-1$, imply SIC$_{i,j}$ for all $i = 1, \ldots, m$ and all $j>i$.
\end{prop}

\noindent \textsc{Proof. } Fix $i$; by hypothesis the local constraints SIC$_{k, k+1}$ hold for all $k = 1, \ldots, m-1$. We establish SIC$_{i,j}$ for all $j > i$ by induction on $j$, upwards from $j = i + 1$.

\noindent Base case $j = i + 1$: This is SIC$_{i, i+1}$.

\noindent Induction hypothesis: For some $j$ with $i + 1 \leq j \leq m - 1$, SIC$_{i,j}$ holds, that is, $t_{i} - \theta_{i} q_i > t_{j} - \theta_{i} q_{j}$.

\noindent Induction step: We show that SIC$_{i, j+1}$ holds. By SIC$_{j, j+1}$,
\begin{align*}
t_{j} - t_{j + 1} > \theta_{j}\left(q_{j} - q_{j + 1}\right).
\end{align*}
By monotonicity $q_{j} > q_{j + 1}$, so that $q_{j} - q_{j + 1} > 0$; since $j > i$ implies $\theta_{j} > \theta_{i}$, it follows that $\theta_{j}\left(q_{j} - q_{j + 1}\right) > \theta_{i}\left(q_{j} - q_{j + 1}\right)$ and hence:
\begin{align*}
t_{j} - t_{j + 1} > \theta_{i}\left(q_{j} - q_{j + 1}\right); \quad \mbox{ that is, } \quad t_{j} - \theta_{i} q_{j} > t_{j + 1} - \theta_{i} q_{j + 1}.
\end{align*}
Combining the last inequality with the induction hypothesis,
\begin{align*}
t_{i} - \theta_{i} q_i > t_{j} - \theta_{i} q_{j} > t_{j + 1} - \theta_{i} q_{j + 1},
\end{align*}
which is SIC$_{i, j+1}$. Upward induction from $j = i + 1$ to $j = m$ therefore yields SIC$_{i,j}$ for all $j > i$. \hfill $\Box$

\begin{prop}\label{LSIC} Assume that $q_i>q_{i + 1}$ for all $i =1,\ldots,m-1$. Then the local constraints SIC$_{k, k-1}$, $k = 2, \ldots, m$, imply SIC$_{i,j}$ for all $i = 1, \ldots, m$ and all $j<i$.
\end{prop}

\noindent \textsc{Proof. } Fix $i$; by hypothesis the local constraints SIC$_{k, k-1}$ hold for all $k = 2, \ldots, m$. We establish SIC$_{i,j}$ for all $j < i$ by induction on $j$, downwards from $j = i - 1$.

\noindent Base case $j = i - 1$: This is SIC$_{i, i-1}$.

\noindent Induction hypothesis: For some $j$ with $2 \leq j \leq i - 1$, SIC$_{i,j}$ holds, that is, $t_{i} - \theta_{i} q_i > t_{j} - \theta_{i} q_{j}$.

\noindent Induction step: We show that SIC$_{i, j-1}$ holds. By SIC$_{j, j-1}$,
\begin{align*}
t_{j} - t_{j - 1} > \theta_{j}\left(q_{j} - q_{j - 1}\right).
\end{align*}
By monotonicity $q_{j - 1} > q_{j}$, so that $q_{j} - q_{j - 1} < 0$; since $j < i$ implies $\theta_{j} < \theta_{i}$, it follows that $\theta_{j}\left(q_{j} - q_{j - 1}\right) > \theta_{i}\left(q_{j} - q_{j - 1}\right)$ and hence:
\begin{align*}
t_{j} - t_{j - 1} > \theta_{i}\left(q_{j} - q_{j - 1}\right); \quad \mbox{ that is, } \quad t_{j} - \theta_{i} q_{j} > t_{j - 1} - \theta_{i} q_{j - 1}.
\end{align*}
Combining the last inequality with the induction hypothesis,
\begin{align*}
t_{i} - \theta_{i} q_i > t_{j} - \theta_{i} q_{j} > t_{j - 1} - \theta_{i} q_{j - 1},
\end{align*}
which is SIC$_{i, j-1}$. Downward induction from $j = i - 1$ to $j = 1$ therefore yields SIC$_{i,j}$ for all $j < i$. \hfill $\Box$\\ 

The previous results allow us to reduce the constraints to the following:
\begin{itemize}
\item[] SIC$_{i,i+1}$ for all $i=1,\ldots,m-1$: $t_{i} - \theta_{i}q_i > t_{i + 1} - \theta_{i}q_{i + 1}$,
\item[] SIC$_{i,i-1}$ for all $i=2,\ldots,m$: $t_{i} - \theta_{i}q_i > t_{i-1} - \theta_{i}q_{i - 1}$,
\item[] SPC$_m$: $t_{m} - \theta_{m} q_{m} > 0$,
\item[] SM$_{i,i+1}$ for all $i=1,\ldots,m-1$: $q_i > q_{i + 1}$.
\end{itemize}
Optimal transfers are such that the relevant SIC and SPC constraints ``bind'': 
\begin{align*} 
t_{m}&=\left\lceil \theta_m\cdot q_{m}\right\rceil+ 1 - \left\lceil \frac{1}{\gamma}\right\rceil,
\end{align*} 
and for $i=1,\ldots,m-1$, by SIC$_{i,i+1}$, 
\begin{align*}
t_{i} &=\left\lceil \theta_{i} q_i +t_{i + 1} - \theta_{i} q_{i + 1} \right\rceil+ 1 - \left\lceil \frac{1}{\gamma}\right\rceil
\\
t_{i} &=\left\lceil t_{i + 1}+\theta_{i}\left(q_i - q_{i + 1}\right) \right\rceil+ 1 - \left\lceil \frac{1}{\gamma}\right\rceil
\\
t_{i} &= t_{i + 1}+\left\lceil\theta_{i}\left(q_i - q_{i + 1}\right) \right\rceil+ 1 - \left\lceil \frac{1}{\gamma}\right\rceil.
\end{align*} 
As explained in the main text, the term $1 - \left\lceil \sfrac{1}{\gamma}\right\rceil$ is nonzero only when $\gamma = \infty$, where it takes the value 1. 

\begin{claim}\label{product} For any $n \in D$ and $j = 1, \ldots, m$, $\left\lceil \left(j - \sfrac{1}{\gamma}\right) n \right\rceil = \left\lceil j - \sfrac{1}{\gamma}\right\rceil n=jn$.
\end{claim}

\noindent \textsc{Proof. } As $\gamma > b$ implies that $\sfrac{n}{\gamma} \in [0, 1)$ for any $n \in D$, we have $\left\lceil \left(j - \sfrac{1}{\gamma}\right) n \right\rceil = \left\lceil j n - \sfrac{n}{\gamma} \right\rceil = jn + \left\lceil - \sfrac{n}{\gamma} \right\rceil = jn$. \hfill $\Box$\\

Notice that, for every $\theta_{j}$ with $j = 1, \ldots, m$, we have that $\lceil \theta_{j}\rceil = j$. Thus, we are able to mimic the usual characterization of optimal transfers.

\begin{prop}\label{binding} Assume that strict monotonicity holds and that no type is shut down, $q_{m} \geq 1$.\footnote{The requirement $q_{m} \geq 1$ binds only for $\gamma < \infty$, where the least efficient type's rent is $t_{m} - \theta_{m} q_{m} = \frac{1}{\gamma}q_{m}$. For $\gamma = \infty$ the extra dollar leaves him a rent of 1 irrespective of $q_{m}$.} Then optimal transfers are $t_{m}=m q_{m}+ 1 - \left\lceil \sfrac{1}{\gamma}\right\rceil$ and, for $i=1,\ldots,m-1$:
\[ 
t_{i} = i q_i + \sum_{j =i+1}^{m} q_{j}+(m-i+1)\left(1-\left\lceil \frac{1}{\gamma}\right\rceil\right).
\]
\end{prop}

\noindent \textsc{Proof. } Most of the argument is laid out in the text. From Claim~\ref{product}, 
\begin{align*}
t_{i} &= t_{i + 1}+\left\lceil\theta_{i}\left(q_i - q_{i + 1}\right) \right\rceil+1-\left\lceil \frac{1}{\gamma}\right\rceil
\\
&= t_{i + 1}+\left\lceil\theta_{i} \right\rceil\left(q_i - q_{i + 1}\right)+1-\left\lceil \frac{1}{\gamma}\right\rceil
\\
&= t_{i + 1}+i \left(q_i - q_{i + 1}\right)+1-\left\lceil \frac{1}{\gamma}\right\rceil
\\
&=i q_i + t_{i + 1}-i q_{i + 1}+1-\left\lceil \frac{1}{\gamma}\right\rceil.
\end{align*} 
By recursive substitution,
\begin{align*}
t_{m-1}&=(m-1)q_{m-1}+t_{m}-(m-1)q_{m}+1-\left\lceil \frac{1}{\gamma}\right\rceil
\\
&=(m-1)q_{m-1}+m q_{m}+1-\left\lceil \frac{1}{\gamma}\right\rceil-(m-1)q_{m}+1-\left\lceil \frac{1}{\gamma}\right\rceil
\\
&=(m-1)q_{m-1}+q_{m}+2\left(1-\left\lceil \frac{1}{\gamma}\right\rceil\right);
\\
t_{m-2}&=(m-2)q_{m-2}+ t_{m-1}-(m-2)q_{m-1}+1-\left\lceil \frac{1}{\gamma}\right\rceil
\\
&=(m-2)q_{m-2}+q_{m-1}+q_{m}+3\left(1-\left\lceil \frac{1}{\gamma}\right\rceil\right);
\end{align*} 
and, for $j=3,\ldots,m-1$,
\begin{align*}
t_{m-j}&=(m-j)q_{m-j}+\sum_{h=0}^{j-1}q_{m-h}+(j+1)\left(1-\left\lceil \frac{1}{\gamma}\right\rceil\right).
\end{align*} 
The proof is complete once we set $i=m-j$. \hfill $\Box$\\

As noted at the beginning of this appendix, Proposition~\ref{binding} is stated for the formulation with strict inequality constraints. For the formulation with weak inequality constraints, simply delete the term $1 - \left\lceil \sfrac{1}{\gamma}\right\rceil$ throughout; the argument remains otherwise unchanged. The two cases are collected in equations~\eqref{optimal_ts_m} and~\eqref{optimal_ts_i} in the main text by means of the indicator $\mathbb{I}_s$.

In standard screening, the binding local upward incentive constraints imply the downward local incentive constraints. The argument under discreteness is laid out below.

\begin{prop}\label{local_SIC} Assume that strict monotonicity holds and that transfers are as in Proposition~\ref{binding}. Then, for all $i=2,\ldots,m$, SIC$_{i,i-1}$ holds whenever $\gamma \neq \infty$; if $\gamma = \infty$, SIC$_{i,i-1}$ holds if and only if $q_{i - 1} \geq q_i + 2$.
\end{prop}

\noindent \textsc{Proof. } Fix $i = 2, \ldots, m$. By Proposition~\ref{binding} and $\lceil \theta_{i-1} \rceil = i - 1$,
\begin{align*}
t_{i-1} - t_{i} = \left(i - 1\right)\left(q_{i - 1} - q_i\right) + 1 - \left\lceil\frac{1}{\gamma}\right\rceil.
\end{align*}
Hence SIC$_{i,i-1}$, that is $t_{i} - \theta_{i} q_i > t_{i-1} - \theta_{i} q_{i - 1}$, is equivalent to:
\begin{align*}
\theta_{i}\left(q_{i - 1} - q_i\right) &> \left(i - 1\right)\left(q_{i - 1} - q_i\right) + 1 - \left\lceil\frac{1}{\gamma}\right\rceil,
\end{align*}
and therefore, since $\theta_{i} = i - \sfrac{1}{\gamma}$, to:
\begin{align*}
\left(1 - \frac{1}{\gamma}\right)\left(q_{i - 1} - q_i\right) &> 1 - \left\lceil\frac{1}{\gamma}\right\rceil.
\end{align*}
For $\gamma \neq \infty$ the right-hand side is $0$ while the left-hand side is positive by strict monotonicity, so SIC$_{i,i-1}$ holds. For $\gamma = \infty$ the condition reads $q_{i - 1} - q_i > 1$, so that SIC$_{i,i-1}$ holds if and only if $q_{i - 1} \geq q_i + 2$. This is a further reason why the formulation with strict inequality constraints requires types that are off-set from the integers; under weak inequality constraints no such difficulty arises, as the corresponding condition becomes $\left(1 - \sfrac{1}{\gamma}\right)\left(q_{i - 1} - q_i\right) \geq 0$. \hfill $\Box$

\appendixnumbering


\section{Log-concave Beliefs}\label{others}

This appendix includes properties of log-concave beliefs that are well known, but are featured here for the sake of completeness and self-containment.

\begin{prop}\label{hazard_rates} Let $p\in\Delta(\Theta)$ be a full-support distribution. If $p$ is log-concave,
\begin{itemize}
\item[(i)] Relative likelihoods are nonincreasing in $i$: For any $i, j = 1, \ldots,m$ and $k$ such that $j > i$ and $j+k \leq m$, $$\frac{p_{i+k}}{p_i} \geq \frac{p_{j+k}}{p_j}.$$
\item[(ii)] The discrete Mills' ratio, $\frac{\sum_{k<i} p_{k}}{p_{i}}$, is strictly increasing in $i$.
\end{itemize}
\end{prop}

\noindent \textsc{Proof. } For $i = 1, \ldots, m-2$,
\begin{eqnarray*} (p_{i + 1})^{2} & \geq & p_i p_{i+2} \\
\frac{p_{i + 1}}{p_i} & \geq & \frac{p_{i+2}}{p_{i + 1}} \\
\log p_{i + 1} - \log p_i & \geq & \log p_{i+2} - \log p_{i + 1}.
\end{eqnarray*}
Inductively, we have for $j > i$, $j+1 \leq m$, $$\log p_{i + 1} - \log p_i \geq \log p_{j+1} - \log p_{j}.$$

(i) For any $i, j = 1, \ldots, m$ and $k$ such that $j > i$ and $j+k \leq m,$
\begin{eqnarray*} \frac{p_{i+k}}{p_i} & = & \frac{p_{i + 1}}{p_i} \frac{p_{i+2}}{p_{i + 1}} \cdots \frac{p_{i+k}}{p_{i+k-1}} \\
\log\left(\frac{p_{i+k}}{p_i}\right) & = & \log\left(\frac{p_{i + 1}}{p_i} \frac{p_{i+2}}{p_{i + 1}} \cdots \frac{p_{i+k}}{p_{i+k-1}}\right) \\
& = & (\log p_{i + 1} - \log p_i) + (\log p_{i+2} - \log p_{i + 1}) + \cdots +  (\log p_{i+k} - \log p_{i+k-1}) \\
& \geq & (\log p_{j+1} - \log p_j) + (\log p_{j+2} - \log p_{j+1}) + \cdots +  (\log p_{j+k} - \log p_{j+k-1}) \\
& = &
\log\left(\frac{p_{j+1}}{p_j} \frac{p_{j+2}}{p_{j+1}} \cdots \frac{p_{j+k}}{p_{j+k-1}}\right) \; = \; \log\left(\frac{p_{j+k}}{p_j}\right),
\end{eqnarray*} where the inequality follows from log-concavity applied to each term of the sum. Since $\log$ is strictly increasing, (i) follows.

(ii) From (i), for $k<i$, we have $p_i p_{k+1}\geq p_{k} p_{i + 1}$. Then, since $p$ has full support,
\begin{eqnarray*} 
p_i\sum_{k=1}^{i-1}p_{k+1} & \geq & p_{i + 1}\sum_{k=1}^{i-1}p_{k} \\
p_i\sum_{k=2}^{i}p_{k} & \geq & p_{i + 1}\sum_{k=1}^{i-1}p_{k} \\
p_i\sum_{k=1}^{i}p_{k} & > & p_{i + 1}\sum_{k=1}^{i-1}p_{k} \\
p_i\sum_{k<i+1}p_{k} & > & p_{i + 1}\sum_{k<i}p_{k}, \\
\end{eqnarray*}
where the third line adds $p_i p_{1}>0$ to the left-hand side. Rearranging establishes the result. \hfill $\Box$

\begin{cor}\label{monotonicity_virtual_valuations} For any $p \in \Delta^{FL}(\Theta)$, $\varphi_{i+1}(p)-\varphi_i(p)>1$---and hence $\varphi_{i+1}(p)>\varphi_i(p)$---for all $i = 1,\ldots, m-1$. 
\end{cor}

\noindent \textsc{Proof. } By the definition of virtual costs, $\varphi_{i+1}(p)-\varphi_i(p) = 1 + \frac{\sum_{k<i+1} p_{k}}{p_{i+1}} - \frac{\sum_{k<i} p_{k}}{p_{i}}$. By part (ii) of Proposition~\ref{hazard_rates}, $\frac{\sum_{k<i+1} p_{k}}{p_{i+1}}>\frac{\sum_{k<i} p_{k}}{p_{i}}$, and thus $\varphi_{i+1}(p)-\varphi_i(p)>1$. \hfill $\Box$\\

We close with the proof of the nonemptiness claim of Section~\ref{sec: screening}.

\noindent \textsc{Proof of Remark~\ref{nonempty}. } If: Let $\Delta^{+} v(0) > m$, fix $\delta > 0$, and for $\alpha \in \mathbb{R}$ let $p^{\alpha}$ be given by $p^{\alpha}_i \propto e^{\alpha i - \delta i^{2}}$, $i = 1, \ldots, m$. Each $p^{\alpha}$ has full support, and the second difference of $\log p^{\alpha}_i$ equals $- 2 \delta < 0$, so $p^{\alpha}$ is strictly log-concave. Moreover, for every $j < m$,
$$\frac{p^{\alpha}_j}{p^{\alpha}_{m}} = e^{\alpha \left(j - m\right) + \delta \left(m^{2} - j^{2}\right)} \longrightarrow 0 \quad \mbox{ as } \alpha \longrightarrow \infty,$$
so that $\varphi_{m}(p^{\alpha}) = m + \sum_{j < m} p^{\alpha}_j/p^{\alpha}_{m}$ converges to $m$. Hence $\varphi_{m}(p^{\alpha}) < \Delta^{+} v(0)$ for $\alpha$ large enough, and for such $\alpha$ we have $p^{\alpha} \in \Delta^{FSLN}(\Theta) \subseteq \Delta^{FLN}(\Theta)$.

Only if: Take any $p \in \Delta^{FLN}(\Theta)$. Since $p$ has full support, $\varphi_{m}(p) = m + \frac{\sum_{j < m} p_j}{p_{m}} > m$, and since $p$ shuts down no type, $\Delta^{+} v(0) > \varphi_{m}(p)$. Hence $\Delta^{+} v(0) > m$. \hfill $\Box$\\

\appendixnumbering


\section{Nonlinear Cost Structures}\label{non_linear_app}

This appendix explores nonlinear cost structures. We assume that the agent's cost is given by a function $c(q,\theta_{i})$ that has the following properties:

\begin{ass}\label{properties_c_basic} The cost function $c(q,\theta_{i})$ satisfies the following properties:
\begin{enumerate}
\item\label{normalization_c}
For every $\theta\in\Theta$, $c(0,\theta)=0$;
\item\label{sincreasing_c}
$c(q,\theta_{i})$ is strictly increasing in $q$;
\item\label{sdc_c}
Discrete convexity in $q$: $c(q + 1,\theta_{i})+c(q - 1,\theta_{i})\geq 2 c(q,\theta_{i})$ for every $q \in \{1, \ldots, b - 1\}$ and $i=1,\ldots,m$;
\item\label{ssuperm}
Strict increasing differences: $c(q+1,\theta_{i+1})-c(q,\theta_{i+1})>c(q+1,\theta_{i})-c(q,\theta_{i})$ for every $q \in \{0, \ldots, b - 1\}$ and $i=1,\ldots,m-1$;
\item\label{id_marginal}
Increasing differences in marginal costs: $c(q + 1,\theta_{i+1})+c(q - 1,\theta_{i+1})-2c(q,\theta_{i+1}) \geq c(q + 1,\theta_{i})+c(q - 1,\theta_{i})-2 c(q,\theta_{i})$ for every $q \in \{1, \ldots, b - 1\}$ and $i=1,\ldots,m-1$.
\end{enumerate}
\end{ass} 

Assumptions~\ref{properties_c_basic}.\ref{sincreasing_c} and~\ref{properties_c_basic}.\ref{sdc_c} say that the cost is increasing in $q$, at a nondecreasing rate. Assumption~\ref{properties_c_basic}.\ref{normalization_c} is a normalization, which implies that the agent earns $u_A(\bm{o}, \theta)=0$ from the outside option. Assumption~\ref{properties_c_basic}.\ref{ssuperm} ranks the types: Summing it over $k=q,\ldots,q'-1$ yields that $c(q',\theta_{i})-c(q,\theta_{i})$ is strictly increasing in $\theta_{i}$ for every $q' > q$, and the case $q=0$ yields, by Assumption~\ref{properties_c_basic}.\ref{normalization_c}, that $c(q',\theta_{i})$ itself is strictly increasing in $\theta_{i}$ for every $q' \geq 1$. Higher types thus have both higher costs and higher incremental costs. Assumption~\ref{properties_c_basic}.\ref{id_marginal} imposes the increasing-differences requirement of Assumption~\ref{properties_c_basic}.\ref{ssuperm} one level up, on marginal costs rather than on costs themselves: $\Delta_{q}^{+} c(q,\theta_{i})$ itself has increasing differences in $(q,\theta_{i})$. It is what keeps the principal's per-type objective discrete concave in Section~\ref{integer}. 

We can restate these assumptions using \textit{discrete partial derivatives}. Take $\Delta_{q}^{-} c(q,\theta_{i}) := c(q,\theta_{i})- c(q-1,\theta_{i})$ for $q>0$ and $\Delta_{q}^{+} c(q,\theta_{i}) := c(q+1,\theta_{i})- c(q,\theta_{i})$ for $q<b$; these are the backward and forward versions of the marginal cost, respectively. Define $\Delta_{\theta}^{-} c(q,\theta_{i}) := c(q,\theta_{i}) - c(q,\theta_{i-1})$ for $i>1$ and $\Delta_{\theta}^{+} c(q,\theta_{i}) := c(q,\theta_{i+1}) - c(q,\theta_{i})$ for $i<m$. Assumption~\ref{properties_c_basic}.\ref{sincreasing_c} becomes $\Delta_{q}^{-} c(q,\theta_{i}),\Delta_{q}^{+} c(q,\theta_{i})>0$, while the ranking of types just derived reads $\Delta_{\theta}^{-} c(q,\theta_{i}),\Delta_{\theta}^{+} c(q,\theta_{i})>0$ for every $q \geq 1$. For Assumption~\ref{properties_c_basic}.\ref{sdc_c}, we have $\Delta_{q}^+ \Delta_{q}^- c(q,\theta_{i})\geq 0$ for every $i=1,\ldots,m$. Finally, Assumption~\ref{properties_c_basic}.\ref{ssuperm} can be stated as $\Delta_{\theta}^+ \Delta_{q}^- c(q,\theta_{i}),\,\Delta_{\theta}^- \Delta_{q}^- c(q,\theta_{i})>0$ and $\Delta_{\theta}^+ \Delta_{q}^+ c(q,\theta_{i}),\,\Delta_{\theta}^- \Delta_{q}^+ c(q,\theta_{i})>0$. 

Optimal transfers are now given by $t_{m}=\left\lceil c(q_{m},\theta_{m})\right\rceil$ and, for any $i=1,\ldots,m-1$,
\[
t_{i}=\sum_{j=i}^{m-1}\left\lceil c(q_{j},\theta_{j})-c(q_{j+1},\theta_{j})\right\rceil+\left\lceil c(q_{m},\theta_{m})\right\rceil.
\]
The ceiling function may prevent us from breaking up the term $\left\lceil c(q_{j},\theta_{j})-c(q_{j+1},\theta_{j})\right\rceil$, and hence from writing the information rents as the sum of the quantities awarded to all higher-cost types and maximizing the principal's expected payoff type by type. 

\subsection{Integer Costs}\label{integer}

An obvious case where we can avert the problem and recover the additively separable structure of the principal's objective is when $c(q,\theta_{i})$ is an integer for all $q\in D$ and all $i=1,\ldots,m$. In this case, 
the principal's objective becomes $\sum_{i=1}^{m}p_i\left[v(q_i)-C(q_i,\theta_{i})\right]$, where:
\[
C(q_i,\theta_{i}):=c(q_i,\theta_{i})+\frac{\sum_{k<i} p_{k}}{p_{i}}\cdot \Delta^{-}_{\theta}c(q_i,\theta_{i}) \quad \mbox{ for } i \geq 2, \qquad C(q_{1},\theta_{1}):=c(q_{1},\theta_{1}).
\]  The discrete F.O.C.s are now:
\begin{equation}\label{FOCs_new}
\Delta^+ v(q_i)-\Delta^{+}_{q}C(q_i,\theta_{i})\leq 0\leq \Delta^- v(q_i)-\Delta^{-}_{q}C(q_i,\theta_{i})
\end{equation}
for all $i=1,\ldots,m$, with no (information-based) distortion at the top. Write $C(q,\theta_{i}) = (1 + \lambda_{i})c(q,\theta_{i}) - \lambda_{i} c(q,\theta_{i-1})$, where $\lambda_{i} := \sfrac{\sum_{k<i} p_{k}}{p_{i}}$ and hence $\lambda_{1} = 0$; then Assumptions~\ref{properties_c_basic}.\ref{sdc_c} and~\ref{properties_c_basic}.\ref{id_marginal} give $\Delta_{q}^{+}\Delta_{q}^{-} C(q,\theta_{i}) \geq \Delta_{q}^{+}\Delta_{q}^{-} c(q,\theta_{i}) \geq 0$. Together with Assumption~\ref{properties_v_basic}.\ref{sdc}, this makes $v - C(\cdot,\theta_{i})$ discrete strictly concave, so that~\eqref{FOCs_new} characterizes the optimal quantities (Appendix~\ref{discrete_c_o}). In the linear benchmark of Section~\ref{sec: screening} this step is immediate: costs enter the per-type objective $v(q_i) - \varphi_{i}(p) q_i$ linearly, so it inherits the strict discrete concavity of $v$ outright. Integer costs restore separability but not concavity; the latter is what Assumption~\ref{properties_c_basic}.\ref{id_marginal} supplies.

\begin{ex}\label{quasilinear_int} 
Consider the utility function $v(q) = 250q - 0.5q^{2}$, the grid $D = \{0, 1, \ldots, 15,000\}$, types $\Theta=\{1-\sfrac{1}{\gamma},2-\sfrac{1}{\gamma}\}$ with $\gamma = 20,000$, and prior $p_{1}=p_{2}=\frac{1}{2}$; we have $\Delta^{+} v(q) = 249.5-q$ and $\Delta^{-} v(q)=250.5-q$. Let the cost function be $c(q,\theta_{i})=i^2\cdot q^{2}$. For $\theta_{1}$, we have $\Delta^{+}_{q}c(q,\theta_{1})=2q+1$ and $\Delta^{-}_{q}c(q,\theta_{1})=2q-1$. So,
\begin{align*}
249.5-q_{1}-(2q_{1}+1)&\leq0\leq250.5-q_{1}-(2q_{1}-1)
\\
248.5-3q_{1}&\leq0\leq251.5-3q_{1}
\\
248.5&\leq3q_{1}\leq251.5,
\end{align*}
which yields $q_{1}=83$. For $\theta_{2}$, we have $C(q,\theta_{2})=7q^{2}$, hence  $\Delta^{+}_{q}C(q_{2},\theta_{2})=7(2q_{2}+1)$ and $\Delta^{-}_{q}C(q_{2},\theta_{2})=7(2q_{2}-1)$. The discrete F.O.C. is:
\begin{align*}
249.5-q_{2}-7(2q_{2}+1)&\leq0\leq250.5-q_{2}-7(2q_{2}-1)
\\
242.5-15q_{2}&\leq0\leq257.5-15q_{2}
\\
242.5&\leq15q_{2}\leq257.5,
\end{align*}
so $q_{2}=17$. \hfill $\Box$
\end{ex}

\subsection{Non-Integer Costs}\label{non_integer}

In order to showcase the possibility of distortions at the top, we start with the first-best benchmark. In the absence of incentive constraints, the principal can set $t_{i}=\left\lceil c(q_i,\theta_{i})\right\rceil$ for all $i=1,\ldots,m$. The first-best allocation rule maximizes: 
\[
\text{{\bf(FB)}}\hspace{4mm}\sum_{i=1}^{m}p_i\left[v(q_i)-\left\lceil c(q_i,\theta_{i})\right\rceil\right].
\]
Under asymmetric information, the principal maximizes:
\begin{equation}\label{payoff_general}
\text{{\bf(SB)}}\hspace{4mm}\sum_{i=1}^{m}p_i\left[v(q_i)-\left(1+\frac{\sum_{k<i} p_{k}}{p_{i}}\right)\left\lceil c(q_i,\theta_{i})-c(q_{i + 1},\theta_{i}) \right\rceil\right],
\end{equation}
with $q_{m+1}:=0$, so that $c(q_{m+1},\theta_{m})=0$ by Assumption~\ref{properties_c_basic}.\ref{normalization_c} (and with the normalization $\sum_{k<1} p_{k}\equiv 0$), subject to the monotonicity constraints.

For every $i=1,\ldots,m-1$, the term in~\eqref{payoff_general} corresponding to $i$ involves $q_{i+1}$---including the term for $i=1$. Thus, the problem becomes a (nonseparable) finite-horizon, discrete dynamic programming problem, with the quantity for the ``next-lower'' type as the state variable. For every $i=2,\ldots,m-1$, the optimal quantity $q_i$ solves:
\begin{align*}\label{Bellman}
\mathbf{V}_{i}(q_{i - 1})=\max_{q_i\in D}\left\{p_{i - 1}\left[v(q_{i - 1})-\left(1+\frac{\sum_{k<i-1} p_{k}}{p_{i-1}}\right)\left\lceil c(q_{i - 1},\theta_{i-1})-c(q_i,\theta_{i-1}) \right\rceil\right]+\mathbf{V}_{i+1}(q_i)\right\},
\end{align*}
with:
\begin{align*}
\mathbf{V}_{m}(q_{m-1}):=\max_{q_m\in D}&\left\{p_{m-1}\left[v(q_{m-1})-\left(1+\frac{\sum_{k<m-1} p_{k}}{p_{m-1}}\right)\left\lceil c(q_{m-1},\theta_{m-1})-c(q_{m},\theta_{m-1}) \right\rceil\right]\right.
\\
&\left.+p_{m}\left[v(q_{m})-\left(1+\frac{\sum_{k<m} p_{k}}{p_{m}}\right)\left\lceil c(q_{m},\theta_{m})\right\rceil\right]\right\}
\end{align*}
and $\mathbf{V}_{1}:= \max_{q_{1}\in D}\mathbf{V}_{2}(q_{1})$.

A small reduction in the cost differential $c(q_i,\theta_{i})-c(q_{i+1},\theta_{i})$ for a given $i$ can lead to saving a full ``dollar'' in transfers. Moreover, by the ``moving average'' structure of transfers, such effects can be chained together to impact the optimal quantities for nonadjacent types. As a result, (weak) monotonicity of the allocation may now break down even under strictly log-concave priors (Example~\ref{monot_fails}), and we may have upward distortions at the top (Example~\ref{quasilinear_nonint}).

\begin{figure}[h!]
\centering
\includegraphics[scale=0.5]{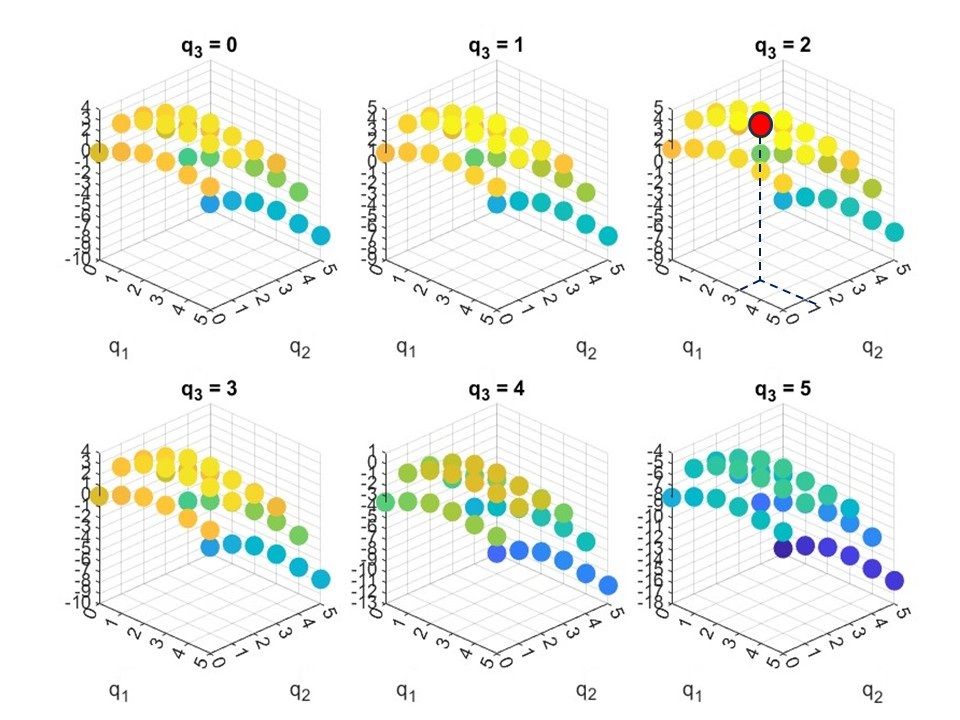}
\caption{Monotonicity can fail even under a log-concave prior; see Example~\ref{monot_fails}.\label{monot_fails_graph}}
\end{figure}

\begin{ex}\label{monot_fails} 
Take $\Theta=\{0.99,1.99,2.99\}$ and $D=\{0,1,\ldots,5\}$; $v(q)$ is $v(0)=0$; $v(1)=5.9089$; $v(2)=9.8115$; $v(3)=11.5163$; $v(4)=12.0747$; and $v(5)=12.1609$; the cost function is $c(q,\theta_{i})=h(i)\cdot q^{2}$, with $h(1)=0.1940$; $h(2)=1$; and $h(3)=1.1$. The prior is $p_{1}=0.2$ and $p_{2}=p_{3}=0.4$, so that $\frac{p_{1}}{p_{2}}=\frac{1}{2}$ and $\frac{p_{1}+p_{2}}{p_{3}}=\frac{3}{2}$. Notice that the prior is strictly log-concave: $p_{2}^{2}=0.16>0.08=p_{1}p_{3}$. However, the unique maximizer of the principal's objective, ignoring the monotonicity constraint, is $q_{1}=3$, $q_{2}=1$, and $q_{3}=2$. See Figure~\ref{monot_fails_graph}. \hfill $\Box$
\end{ex}

\begin{ex}\label{quasilinear_nonint} 
Consider the same setting as in Example~\ref{quasilinear_int} except for the cost function, which is now  $c(q,\theta_{i})=49.1\cdot i\cdot q$. The principal maximizes:
\begin{align*}
&p_{1}\left[v(q_{1})-\left\lceil c(q_{1},\theta_{1})-c(q_{2},\theta_{1})\right\rceil \right]+p_{2}\left[v(q_{2})-\left(1+\frac{p_{1}}{p_{2}}\right)\left\lceil c(q_{2},\theta_2)\right\rceil \right]
\\
&=0.5\left[v(q_{1})-\left\lceil 49.1(q_{1}-q_{2})\right\rceil \right]+0.5\left[v(q_{2})-2\left\lceil 98.2 q_{2}\right\rceil \right].
\end{align*}
For $\theta_{1}$, the discrete F.O.C.s are:
\begin{align*}
&249.5-q_{1}-\left(\left\lceil 49.1(q_{1}+1-q_{2})\right\rceil-\left\lceil 49.1(q_{1}-q_{2})\right\rceil\right)
\\
&\leq0\leq250.5-q_{1}-\left(\left\lceil 49.1(q_{1}-q_{2})\right\rceil-\left\lceil 49.1(q_{1}-1-q_{2})\right\rceil\right);
\end{align*}
for $\theta_{2}$,
\begin{align*}
&-\left(\left\lceil 49.1(q_{1}-q_{2}-1)\right\rceil-\left\lceil 49.1(q_{1}-q_{2})\right\rceil\right)+249.5-q_{2}-2\left(\left\lceil 98.2(q_{2}+1)\right\rceil-\left\lceil 98.2q_{2}\right\rceil\right)
\\
&\leq0\leq-\left(\left\lceil 49.1(q_{1}-q_{2})\right\rceil-\left\lceil 49.1(q_{1}-q_{2}+1)\right\rceil\right)+250.5-q_{2}-2\left(\left\lceil 98.2q_{2}\right\rceil-\left\lceil 98.2(q_{2}-1)\right\rceil\right).
\end{align*}
These inequalities have a unique integer solution, $q_{1}=201$ and $q_{2}=103$. The (unique) first-best quantity for the lowest-cost type, however, is 200:\footnote{For $q = 199$, the inequality becomes $200.5\leq 199\leq 201.5$, which is absurd. Similarly, for $q = 201$, we get $200.5\leq 201\leq 200.5$.}
\begin{align*}
249.5-\left(\left\lceil 49.1(q_{1}^{FB}+1)\right\rceil -\left\lceil 49.1q_{1}^{FB}\right\rceil\right)&\leq q_{1}^{FB} \leq 250.5-\left(\left\lceil 49.1q_{1}^{FB}\right\rceil -\left\lceil 49.1(q_{1}^{FB}-1)\right\rceil\right)
\\
249.5-\left(9,870-9,820\right)&\leq 200 \leq 250.5-\left(9,820-9,771\right)
\\
249.5-50&\leq 200 \leq 250.5-49
\\
199.5&\leq200\leq 201.5\hspace{3mm}\checkmark
\end{align*}
Thus, there is an upward distortion at the top. \hfill $\Box$
\end{ex}

In the linear-cost case, a downward distortion at the top is possible because of the rounding up of the types. However, upward distortions are impossible there.


\end{document}